\documentclass[twocolumn]{aastex701}

\graphicspath{{./}{}}
\usepackage{graphicx}
\usepackage{multirow}
\usepackage{adjustbox}
\usepackage{booktabs}
\usepackage{threeparttable}
\usepackage[table]{xcolor}
\usepackage{amsmath}
\usepackage{xcolor}

\begin{document}

\title{The Self-Limiting Nature of Jet-Modulated Thermal Conduction in Cool Core Clusters}

\author{Jennifer N. Stafford}
\affiliation{Department of Astronomy, University of Wisconsin-Madison, 475 N. Charter St, Madison, WI 53706, USA}
\email[show]{jstafford2@wisc.edu}

\author{Sebastian Heinz}
\affiliation{Department of Astronomy, University of Wisconsin-Madison, 475 N. Charter St, Madison, WI 53706, USA}
\email{sheinz@wisc.edu}

\author{Mateusz Ruszkowski}
\affiliation{Department of Astronomy, University of Michigan, 1085 S. University
Ann Arbor, MI 48109-1107, USA}
\email{mateuszr@umich.edu}

\author{Torsten En{\ss}lin}
\affiliation{Max-Planck-Institut f\"ur Astrophysik (MPA), Max-Planck Karl-Schwarzschild-Str.~1 85748 Garching,Germany}
\email{ensslin@mpa-garching.mpg.de}

\author{Yi-Hao Chen}
\affiliation{Department of Astronomy, University of Wisconsin-Madison, 475 N. Charter St, Madison, WI 53706, USA}
\email{ychen@astro.wisc.edu}

\begin{abstract}
Conduction as a mechanism for explaining the disrupted cooling-flow in galaxy clusters has been mostly discounted, as the process is inefficient at transporting heat all the way from the cluster into the core. However, thermal conduction can be strongly enhanced when materials of significantly different temperature are brought into proximity, and thus into close thermal contact. Jets of active galactic nuclei may act as heat pumps by bringing low-entropy gas from the cluster core into thermal contact with the hot outer atmosphere of the cluster, significantly increasing the feedback efficiency of active galactic nuclei. We test this hypothesis by running a suite of 3D magnetohydrodynamic simulations of active galactic nuclei jets in a Perseus-like cluster, including anisotropic conduction. We find that the heat pump efficiency $\eta$ can reach up to 50\% of the maximum possible efficiency $\eta_{\rm\,max}$ if  conduction operates near the Spitzer-Braginskii limit, while $\eta\approx\,f_{\rm\,sp}\eta_{\rm\,max}$ if conduction along the field lines is substantially suppressed below the Spitzer-Braginskii value by a factor $f_{\rm\,sp}$ by kinetic effects, as recently suggested. We further find that jet-induced thermal conduction is self-limiting: Magnetic draping during the uplift results in a magnetic field orientation close to perpendicular to the induced temperature gradients, significantly reducing conduction along the ideal conductive pathways.  Thus, for conservative assumptions about thermal conduction suppression by $f_{\rm\,sp}\,\lesssim\,0.1$, the heat pump effect leads to only marginal heat transfer and, correspondingly, to immaterial changes in the overall thermal evolution of cool core clusters beyond the isolated effects of conduction and jet-induced heating alone.
\end{abstract}

\section{Introduction}
\label{sec:intro}

Galaxy clusters are some of the largest gravitationally bound structures in the universe \citep{Zwicky1933,Zwicky1937}.
Modern nomenclature distinguishes between cool core and non-cool core clusters \citep{bahcall1977clusters,Dressler1984}. This distinction is based on their gas properties derived from X-ray observations, where cool core clusters are generally understood to be dynamically relaxed, while non-cool core clusters are thought to have undergone recent dynamical interaction and have not had enough time to sufficiently cool and relax \citep{Hudson2010}.

\subsection{The Cooling Flow Problem}

The intracluster medium (ICM) cools through thermal X-ray emission. In the absence of heating, the radiative energy loss in cool core clusters would lead to catastrophic, runaway cooling in the form of a classical cooling flow on timescales substantially shorter than a Hubble time, resulting in the accumulation of cold gas and elevated star formation rates \citep{Fabian_94, allen97}. The absence of the expected rate of star formation and the lack of cold gas observed in clusters has given rise to the well-known {\em cooling flow problem}.

The advent of the modern X-ray observatories \textit{Chandra X-ray} and \textit{XMM-Newton} gave astronomers the ability to spatially and spectrally resolve the intracluster medium (ICM) 
\citep{McNamara_Nulsen07,Fabian2012}, revealing that cool core clusters are deficient in cooling gas at or below $\sim1$–$2$~keV. This suggests they are not following the expected cooling flow, challenging our understanding of cooling models \citep{Peterson_03, Peterson_06}. This became known as the second \textit{cooling flow problem} (the first being the discovery of X-ray cooling flows without an equivalent amount of star formation; \citet{Fabian1977}).

The solution to the cooling flow problem is generally thought to be the injection of energy from jets of active galactic nuclei (AGN) launched by the central brightest cluster galaxies (BCGs)\citep{Morgan1958,Morgan1969,bautz1970classification,schneider1983surface}. These central galaxies often host radiatively inefficient AGN which can launch powerful ($\gtrsim10^{45}$~erg/sec) relativistic jets which can extend well beyond their host galaxy, reaching extents of over 100~kpc into the cluster atmosphere \citep{Birzan2004,McNamara_Nulsen07,Fabian2012}.

While the microphysical details are still uncertain, the prevalence of radio galaxies and X-ray cavities in the centers of cool core clusters suggests that active galactic nuclei are a key component in generating and regulating the required heat input to balance the cooling flow \citep{mcnamara00,Birzan2004,mcnamara05,McNamara_Nulsen07,fabian02}.

Although AGN feedback, in general, remains the leading explanation \citep{Fabian1999}, other processes have been proposed to offset cooling in cluster cores. For example, cluster-scale effects such as sloshing of gas induced by mergers may contribute to the heating, but they are generally inefficient and not coupled causally to the thermal state of the gas \citep{ZuHone2010}. 

Initially, thermal conduction between the cool core and the hot cluster atmosphere had also been considered as a potential heating mechanism. However, thermal conduction is mostly discounted as a feedback mechanism for three main reasons: the low efficiency across large distances, the strong temperature dependence of the conduction coefficient limiting conductive heating of the coldest, most rapidly cooling gas that may give way to runaway cooling, and the significantly reduced effective anisotropic conduction rates in the magnetized ICM\citep{Zakamska_2003,Voigt2004,Dolag2004}. Additionally, recent work suggests that conduction may be suppressed even {\em along} the field \citep{Roberg-Clark_2016,Roberg_Clark_2018}, further reducing its efficiency. AGN feedback thus remains the most compelling mechanism for regulating cooling in cluster cores \citep{Ciotti1997,Silk1998,Fabian1999,Burns1990,Pedlar1990,Baum1991}.

\subsection{The Heat Pump Model}
Magnetohydrodynamic (MHD) simulations are the primary tool used to investigate the impact of AGN on cluster environments. MHD simulations allow for a more physically accurate study of the interactions of gas in a magnetized medium
than purely hydrodynamic models through the inclusion of magnetic fields.
This is particularly important, as observations of Faraday rotation and synchrotron polarization have revealed the presence of magnetic fields in galaxy clusters \citep{Carilli2002,Govoni2004}. Full MHD modeling of cluster environments is relatively recent due to the computational power needed to evolve magnetic fields alongside gas dynamics and additional microphysics such as conduction \citep{Ruszkowski2007,Dolag_2009,Yang2016,Ehlert_2021}.

Early studies explored the ability of jets to inflate bubbles and redistribute energy within the ICM, showing that jet-driven structures could plausibly disrupt cooling and transport energy outward \citep[e.g.,][]{vernaleo2006,heinz2006,Morsony2010, Gaspari2011}. Other studies have also looked at jet-inflated lobe morphology, magnetic field topology, and radio observables \citep[e.g.,][]{Hardcastle2014, Weinberger2017}, though they generally do not address thermal conduction or its role in AGN feedback. More recent works have incorporated increasingly complex physics, including turbulence, cosmic rays, and limited magnetic field effects \citep[e.g.,][]{Li2015,Yang2016, Ruszkowski2017, Ehlert_2021}. However, these models have also generally neglected thermal conduction or assumed simplified conduction prescriptions, as it was not considered important, and was computationally expensive, limiting their ability to probe how conduction and AGN feedback interact in detail.

One of the outstanding challenges in radio mode feedback in clusters is a detailed understanding of how the energy injected by AGN is thermalized in the ICM. This presents some challenges, as heating should be sufficiently gentle to leave the cluster gas stratified, rather than shock heating a small volume of gas. 

A pioneering study by \citet{Karen_Yang_2016} looked at anisotropic conduction and cooling in cool core clusters and found conductive heating could offset around 10~\% of the radiative cooling losses.
This motivated long term time evolution studies with higher resolution to further test how conduction could play a role in the heating of these cluster cores \citep{Chen_2019}.

Based on a set of numerical simulations studying feedback, \citet{Chen_2019} pointed out that the motions driven by AGN jets and jet-inflated cavities transport cold central gas from the cluster center into thermal contact with the hotter outer atmosphere such that thermal conduction can be strongly enhanced.
This is analogous to the principle behind geothermal heat pumps. 
Such uplift is observed both in simulations and X-ray temperature maps of clusters\citep[e.g.][]{belsole:2001,molendi:2002,kirkpatrick:2011,gitti:2011}

In their proof-of-concept study, \citet{Chen_2019} argued that the heat flux in clusters may be fast enough to add substantially to the overall heating efficiency of AGN jets even if conduction is substantially suppressed below the canonical isotropic Spitzer-Braginskii value---the maximum uninhibited energy transfer rate \citep{spitzer1962physics,Braginskii1965}.
The work by \citet{Chen_2019} showed that, in principle, the net heating efficiency by AGN jets can exceed 100\%, provided that conduction is sufficiently fast to occur during a dynamical time. 

\setcounter{footnote}{0}

\citet{Chen_2019} provided a proof of concept with promising ramifications; however, they did not actually include the effects of a magnetized intracluster medium or thermal conduction. All calculations were performed in post-processing. Other recent simulation work also explored AGN-driven transport and conduction, incorporating increasingly realistic physics \citep[e.g.][]{Yang2016, Kannan_2016, Kannan_2017, Li2017, Ehlert_2018, barnes2019}. However, most of these models, often due to computational limits, did not self-consistently include both ICM magnetic fields and anisotropic conduction, leaving the full picture incomplete, calling for an ab-initio investigation that includes fully self-consistent models of thermal conduction in the presence of AGN jet feedback. 

Building on the \citet{Chen_2019} study, we implement the missing physics to begin testing whether jet-induced thermal conduction can indeed play a significant role in thermal evolution of cool core clusters.

The structure of this paper is as follows: In Section \ref{sec:methods}, we describe a suite of full 3D MHD cluster simulations including AGN jets and describe the setup and parameters used in our model, along with updates to the \citet{Chen_2019} framework. Section \ref{sec:results} presents the results of our study and outline the analysis techniques used to draw our conclusions. This is followed in Section~\ref{sec:discussion} by a discussion of our findings and their limitations, including key caveats and opportunities for future studies and follow-up. We present a brief convergence discussion is \S\ref{sec:appendix}.

\section{Numerical Methods}
\label{sec:methods}

\subsection{Simulation Setup}
\label{sec:sim}

The simulations by \citet{Chen_2019} were performed using the \texttt{FLASH} code\footnote{\url{https://flash.rochester.edu}} \citep[e.g.][]{Fryxell2000,Dubey2009} to simulate jet feedback aimed specifically at understanding the heating and cooling mechanisms in the Perseus cluster.  Most of the data output is analyzed using the \texttt{Python}-based software package \texttt{yt}\footnote{\url{https://yt-project.org}} \citep{Turk_2011}. For reference below, we refer to our fiducial run as the jet simulation with high conduction (JHC) that includes both jet feedback and anisotropic conduction, highlighted in yellow in Tab.~\ref{tab:simulation-variables}.

As we intend to investigate the feasibility of the heat pump model, we adapt the simulation setup used by \cite{Chen_2019}, extended to include the additional physics necessary to quantify the efficiency of the heat pump. Thus, we use the same \texttt{FLASH} code \citep{Fryxell2000, Dubey2009}, version 4.6.2, to perform full 3D MHD simulations of a galaxy cluster with a central AGN jet. Like \citet{Chen_2019,Chen_2023}, we use the unsplit staggered mesh (\texttt{+usm}) solver with the hybrid Riemann solver in third order and the MC slope limiter, with a default value of 0.3 for the Courant–Friedrichs–Lewy (CFL) number. The initial setup was developed to investigate an array of jet and cluster properties, though it did not include conduction. Additionally, the ICM in those simulations was unmagnetized, however, the jet itself was modeled as a full MHD flow with hoop stress collimation (the jet is injected with a resolution of 16 cells across the jet, following the setup in \citet{Chen_2023}). The conduction efficiency was then estimated in post-processing and showed a possibility of exceeding 100\% under optimistic assumptions. We aim to develop this model further and test it with a set of experiments tailored to answer the question of whether conductive heating of the cool core can be enhanced by the action of jets.

The simulations are initialized with a domain size of a 2~x~2~x~2~Mpc cube. The computational grid is in Cartesian coordinates, with periodic boundary conditions in all directions to ensure numerical stability. The box size is chosen to be sufficiently large to eliminate boundary effects, following the same setup as \citet{Chen_2019}.

We choose a maximum adaptive mesh refinement (AMR) level of 14 that allows us to resolve the core down to a minimum of 30~pc resolution at the finest resolution level. The minimum refinement level at the outermost radii of the cluster is 5, giving us 15~kpc resolution. The refinement criteria are based on the second derivatives of density and pressure, as well as momentum flux of the jet fluid to ensure that the jet itself is always resolved at maximum refinement. At late times, after the active jet episode concludes, we reduce the maximum refinement level by 2, i.e., the minimum cell size increases by a factor of four. We further enforce a radial refinement scheme that maintains a staggered minimum refinement level on radial shells, such that the minimum refinement level within a radius of 15 kpc is 9, corresponding to a cell size of 0.96~kpc, and reduces by one for each factor of 1.5 in radius. With this level of refinement, our fiducial run required approximately 350,000 CPU hours of run time. As in \citet{Chen_2019}, we include two scalar tracer fluids for the mass fraction of the ICM and the non-thermal jet plasma.

Most simulations were run for a duration of $\sim$1024~Myrs. In accordance with \citet{Chen_2019,Chen_2023}, the jet was active for the first 10~Myrs and then switched off to investigate the effect of a single episode of jet activity, resulting in an effective duty cycle (time jet is active vs. inactive) of approximately 1\% over the entire simulation. This numerical experiment allows us to isolate the physical mechanisms of uplift and conduction from the complex history of AGN accretion and feedback. It is important to point out that the models presented here are not meant to provide end-to-end simulations of fully self-consistent cluster evolution. Rather, they are meant to investigate the physical processes relevant to the heat pump mechanism in the same way that the simulations by \citet{Chen_2019} investigated the effects of a single AGN episode in isolation.
This approach allows for sufficient time for the gas to be lifted up, conduct heat, and cycle buoyantly as it starts to settle, as proposed by the model. 

Following \citet{Chen_2019}, the cluster is set up as a beta model atmosphere:
\begin{equation}
    \rho_{\rm ICM}(R)=\rho_{\rm core}\left[1 + \left(\frac{R}{R_{\rm core}}\right)^{2}\right]^{\frac{-3\beta}{2}}
\end{equation}
with $\rho_{\rm core}=9.6\times 10^{-26}\,{\rm g/cm^{3}}$, $R_{\rm core}=26.2\,{\rm kpc}$, and $\beta=0.53$, following the best-fit deprojection in \citet{Zhuravleva_2014,Zhuravleva_2015}, as well as the temperature profile from the same fit to the X-ray data of the Perseus cluster,
\begin{equation}
    \label{eq:temperature}
    T(R)=\frac{T_{\rm out}\left[1 + \left(R/R_{\rm core,T}\right)^{3}\right]}{{T_{\rm out}}/{T_{\rm core}} + \left(R/R_{\rm core,T}\right)^{3}}
\end{equation}
with $R_{\rm core,T}=60.0\,{\rm kpc}$, $T_{\rm core}=3.5\,{\rm keV}$, and $T_{\rm out}=7.4\,{\rm keV}$. 

Because we extend the computational domain beyond the box size of \citet{Chen_2019}, we further impose a density floor of $\rho_{\rm min} = 3.3\times10^{-31}\,{\rm g/cm^3}$ (corresponding roughly to the cosmic mean density) to the beta model derived by \citet{Zhuravleva_2014,Zhuravleva_2015}, keeping the temperature constant. While this is not material for the evolution of the inner cluster, it improves stability of the boundary conditions. Hydrostatic boundary conditions in the setup used in \citet{Chen_2019} had proven subject to some numerical instabilities at late times of the simulation.

We assume a cluster potential such that a pure thermal atmosphere of the above makeup is in hydrostatic equilibrium. The corresponding gravitational acceleration derived from this potential is kept constant throughout the simulation; that is, we do not include self-gravity and assume the potential is dominated by a stationary dark matter halo.

For a full description of the implementation of the jet code, including nozzle injection, further refinement criteria, and cluster properties see \citet{Chen_2019,Chen_2023}. 

\subsection{Implementation of Additional Physics}
\label{sec:physics}

In addition to reproducing the numerical setup in \citet{Chen_2019}, we add some important physics to our implementation of the \texttt{FLASH} code.

\subsubsection{Magnetic Field Setup}
\label{sec:bfields}

The implementation of a magnetized intracluster medium (ICM) is necessary to model the effects of anisotropic conduction. To this end, we construct a randomized tangled magnetic field that we initialize the cluster with. We roughly follow the method used in \citet{Ruszkowski2007} to accomplish this: We construct the vector potential $\vec{A}_{\rm rand}$ from a fully random-phase vector field in Fourier space that follows a prescribed power spectrum $P(k)=P_{0}k^{-\alpha}e^{-k/k_{1}}e^{-k_{2}/k}$ with a power spectral index of $\alpha$ and upper and lower cutoffs $k_{1}$ and $k_{2}$, respectively. The coherence length of the field is then given by $\lambda_{\rm B}\sim 2\pi/k_{2}$. For the simulations discussed in this paper, we choose $\alpha=2.5$, following \citep{Ruszkowski2007}, a coherence length of 50~kpc, and short-wavelength cutoff of $\lambda_{min}=2.5\,{\rm kpc}$.

The resulting field is constructed on a uniform grid and then mapped to the AMR block structure of the initial conditions. Accordingly, we perform cubic interpolation of the vector potential on scales smaller than half the minimum wavelength sampled by the initial vector potential grid. To construct the vector potential on a uniform grid, we impose periodicity on the random phase $\vec{A}_{\rm rand}$ at scales larger than 250~kpc, ensuring that the setup fits within available memory. The region of interest in our simulation is smaller than the scale on which the unattenuated vector potential $\vec{A}_{\rm rand}$ is periodic.

\begin{figure*}[h!t]
  \centering
  \includegraphics[width=0.98\textwidth]{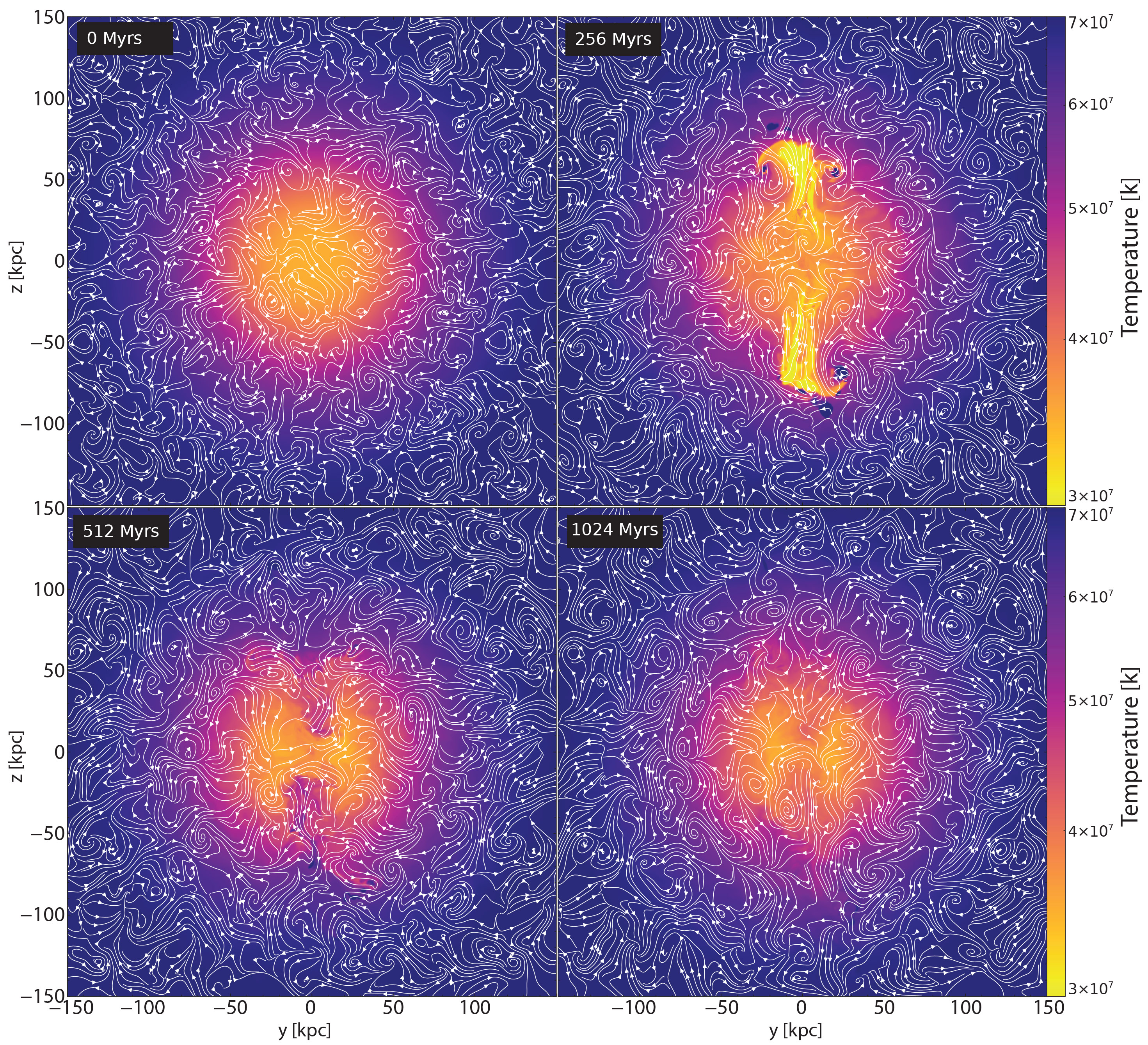}
  \caption{Magnetic field overlay on temperature slices for the 0, 256, 512, and 1024~Myr slices in Fig.~\ref{fig:tempslices}. Panel 1 shows the initial magnetic field setup. Panels 2, 3, and 4 demonstrate how the field lines drape around the uplifted gas, turning off anisotropic conduction by turning the field perpendicular to the temperature gradient between the cooler uplifted gas and the hotter cluster gas.}
  \label{fig:bfieldslices}
\end{figure*}

In order to maintain constant plasma beta $\beta_{\rm P}=P_{\rm gas}/P_{\rm B}=100$, we then employ a radial tapering function $g_{A}=\sqrt{P_{\rm gas}(R)/P_{gas}(R=0)}$, such that the magnetic field pressure follows the thermal pressure profile. We multiply the vector potential by $g_{A}$ such that $\vec{A}=g_{A}\vec{A}_{\rm rand}$ and then construct the divergence-free magnetic field by taking the curl of the resulting vector potential. The normalization of the vector potential $P_{0}$ is set to match the target value of the plasma beta, $\beta_{\rm plasma}=100$, for our fiducial simulation. This guarantees a divergence free magnetic setup. The resulting initial magnetic field setup is visualized in the top left panel of Fig.~\ref{fig:bfieldslices}), and then evolves with the simulation, as visualized in the other panels.

\subsubsection{Anisotropic Conduction}
\label{sec:aniso}

By default, the thermal conduction module in FLASH calculates the isotropic heat flux using the classic Spitzer-Braginskii value for the conduction coefficient,
\begin{equation}
\label{eq:spitzer}
\begin{split}
    \kappa_{sp}  &\sim \,v_{\rm e}\,\lambda_{\rm mfp}\\
                 &\sim 2 \times 10^{30}\,\mathrm{cm^{2}\,s^{-1}}\,\left(\frac{T}{3\,{\rm keV}}\right)^{5/2}\frac{10^{-2}\,{\rm cm^{-3}}}{n_{\rm e}}
\end{split}
\end{equation}
where $v_{\rm e}$ is the thermal velocity of the electrons, $\lambda_{\rm mfp}$ is Coulomb mean free path, and $T$ and $n_{\rm e}$ are the ICM temperature and electron density, respectively.

However, heat conduction in the magnetized intracluster medium is anisotropic, which can heavily suppress the rate of conduction and of conductive heating. This is due to the inability to conduct heat across magnetic field lines as electrons' gyro motion ties them to individual field lines. We deploy an anisotropic conduction module written for \texttt{FLASH} from \cite{Ruszkowski_2011}, which builds upon the isotropic \texttt{SPITZER} conduction module that is part of the regular \texttt{FLASH} distribution.
The anisotropic conduction module uses a monotonized central limiter for conductive fluxes 
\citep{Sharma2007} to ensure stable behavior and avoid negative temperatures when strong temperature gradients are present.

We expect the thermal conduction rate between the non-thermal plasma injected by the jet and the ICM to be insignificant, given that relativistic electrons do not couple well with thermal gas, as Coulomb collisions are strongly suppressed\footnote{Generally, cosmic rays are understood to contribute to heating of the thermal ICM through kinetic effects, however, cosmic ray heating is not the focus of this study and we ignore it.}. Thus, the evolution of the non-thermal plasma is not well represented by the thermal conduction model implemented in \texttt{FLASH}. In fact, the non-thermal plasma injected by the jet presents a potential contamination of the heat conduction signal; therefore, we impose a threshold on the conduction rate based on the presence of non-thermal tracer fluid in a computational cell, as well as an upper limit on the temperature in a given cell, in order to eliminate artificially large conduction from the non-thermal plasma into the uplifted low entropy thermal gas. The threshold is conservatively set to a mass fraction of $f_{\rm jet} \leq 10^{-5}$ and a temperature threshold of $T\leq 8\times 10^{8}\,{\rm K}$, which maintains conduction within the thermal gas but effectively decouples the non-thermal (jet) plasma from our conduction calculations.

\setlength{\tabcolsep}{10pt}   

\begin{deluxetable*}{lccccccc}[t]
\label{tab:simulation-variables}
\tablehead{
   \colhead{\textbf{Run}} & 
   \colhead{\textbf{Power}} & 
   \colhead{\textbf{Plasma Beta}} & 
   \colhead{\textbf{Lambda}} & 
   \colhead{\textbf{Spitzer}} & 
   \colhead{\textbf{Velocity}} & 
   \colhead{\textbf{Jet Duration}} & 
   \colhead{\textbf{Conduction}} \\
   \colhead{} &
   \colhead{(erg s$^{-1}$)} &\colhead{($P_{\mathrm{rad}}/P_{\mathrm{mag}}$)} &
   \colhead{(kpc)} &
   \colhead{Fraction} &
   \colhead{(c)} &
   \colhead{(Myr)} &
   \colhead{}
}

\startdata
\rowcolor{gray!30} C19 & $1.00 \times 10^{45}$ & * & * & * & $0.10$ & $10.00$ & None \\
\rowcolor{gray!30} C19E & $1.00 \times 10^{45}$ & 1000000.00 & * & * & $0.20$ & $10.00$ & None \\
\hline
JNC  & $1.00 \times 10^{45}$ & $100.00$ & $50.00$ & $0.00$ & $0.20$ & $10.00$ & None \\
JLC   & $1.00 \times 10^{45}$ & $100.00$ & $50.00$ & $0.01$ & $0.20$ & $10.00$ & Anisotropic \\
\rowcolor{yellow}JHC   & $1.00 \times 10^{45}$ & $100.00$ & $50.00$ & $0.10$ & $0.20$ & $10.00$ & Anisotropic \\
JHIC   & $1.00 \times 10^{45}$ & $100.00$ & $50.00$ & $0.10$ & $0.20$ & $10.00$ & Isotropic \\
JSC   & $1.00 \times 10^{45}$ & $100.00$ & $50.00$ & $1.00$ & $0.20$ & $10.00$ & Anisotropic \\
\hline
NJNC & * & $100.00$ & $50.00$ & $0.00$ & * & * & None \\
NJLC   & * & $100.00$ & $50.00$ & $0.01$ & * & * & Anisotropic \\
NJHC  & * & $100.00$ & $50.00$ & $0.10$ & * & * & Anisotropic \\
NJHIC  & * & $100.00$ & $50.00$ & $0.10$ & * & * & Isotropic \\
NJSC   & * & $100.00$ & $50.00$ & $1.00$ & * & * & Anisotropic \\
\enddata

\tablecomments{Run name acronyms: NJNC = No Jet, No Conduction; NJLC = No Jet with Low Conduction; NJHC = No Jet with High Conduction; NJHIC = No Jet with High Isotropic Concution at $f_{\rm sp}=0.1$; NJSC = No Jet with Spitzer-Braginskii Conduction at $f_{\rm sp}=1$; JNC = Jet, No Conduction; JLC = Jet with Low Conduction at $f_{\rm sp}=0.01$; JHC = Jet with High Conduction at $f_{\rm sp}=0.1$; JHIC = Jet with High Isotropic Conduction at $f_{\rm sp}$; JSC = Jet with Spitzer-Braginskii Conduction at $f_{\rm sp}=1$; NJIC = No Jet with Isotropic Conduction at $f_{\rm sp}=0.1$. The gray highlighted row (C19) corresponds to \citet{Chen_2019} parameters, and the C19E is our equivalent comparison to the C19 model. We refer to the JHC simulation (highlighted in yellow) as our fiducial run throughout the paper.}

\caption{List of simulations and parameters for the suite of simulations referenced in this paper. The gray highlighted row shows the parameters from the initial \citet{Chen_2019} paper, prior to the addition of conduction and magnetic fields. The top section lists the jet model runs and the bottom section lists the no jet control models.}
\end{deluxetable*} 

As pointed out by \citet{Roberg-Clark_2016,Roberg_Clark_2018}, and as discussed in \citet{Chen_2019}, kinetic effects can substantially suppress conduction even {\em along} the magnetic field lines. X-ray observations of ICM turbulence and discontinuities also suggest that microscopic transport coefficients are suppressed below the Spitzer-Braginskii values \citep{Zhuravleva_2019}. Since the exact magnitude by which conduction is suppressed is currently still under investigation, we introduce a heuristic suppression factor $f_{\rm sp}$ below the classical Spitzer-Braginskii value $\kappa_{\rm sp}$, following \citet{Chen_2019}, such that $\kappa=f_{\rm sp}\kappa_{\rm sp}$. A value of $f_{\rm sp}=1.0$ is the maximum conduction rate plausible in the simulation.

For anisotropic conduction, the heat flux is then given by:
\begin{equation}
    \vec{\mathcal F}_{H}=\kappa_{\rm sp}f_{\rm sp}\nabla\left.T\right|_{\vec B}=\kappa_{\rm sp}f_{\rm sp}\left(\vec{B}\cdot\nabla T\right)\frac{\vec{B}}{B^2}
\end{equation}
and the conductive heating rate is given by $\nabla\cdot\vec{\mathcal F}_{H}$.

We simulate a range of conduction suppression factors, $\mathrm{f}_{\mathrm{sp}}=(0, \, 0.01, \,  0.1, \, 1.0)$, to test the effects of various degrees of suppression, where a value of $f_{\rm sp}=0.01$ correspond to a large degree of microphysical suppression (we thus tag these simulations as ``low conduction''), while a suppression factor of $f_{\rm sp}=0.1$ presents a moderate degree of suppression (we therefore tag these simulations as ``high conduction''). Naturally, a fraction of $f_{\rm sp}=0$ corresponds to a control simulation without any conduction.

In order to evaluate the efficiency of the heat pump effect, we implement scalar variables that track the instantaneous conductive heating rate $\dot{H}=\nabla\cdot \vec{\mathcal F}_{H}$ for each cell, as well as the time integrated total amount of heat $H=\int dt \dot{H}$ transferred into or out of the cell by conduction throughout the simulation. By construction, the total sum of $H$ over the simulation box must vanish. This allows us to isolate the effects of conduction on the thermal evolution of the cluster from all other effects (such as adiabatic changes in temperature, numerical dissipation, or shocks).

\subsection{Computing Paradigm}
\label{sec:computing}

The simulations presented here are designed to perform numerical experiments, with the goal of answering the question of whether jet-modulated conduction can increase the overall feedback efficiency of jets in cool core clusters. As such, they are designed to isolate the relevant physics and do not include competing effects that would make analysis of the experimental outcomes more ambiguous.

Specifically, we do not include externally driven turbulence (e.g., mergers or sloshing), self-gravity (appropriate in an idealized isolated cluster dominated by dark matter), radiative cooling, cosmic ray physics, and sub-grid galaxy physics such as star formation. This implies that we cannot use the simulations to accurately simulate the long-term evolution of clusters, as radiative cooling is fundamental to the evolution of cool core clusters. However, this means we can cleanly separate out the effects of conduction with and without jet feedback, which is the purpose of this paper. This approach also allows us to directly compare our results to the work of \citet{Chen_2019}.

To investigate the effects of AGN jets and conduction within the context of the heat pump mechanism, we designed this suite of models with a set of control and experimental simulations to disentangle the effects of conduction from the effects of the entropy transport by the jet itself. The goal is to determine whether there is an additional conductive heating boost induced by the jet in a magnetized ICM. The control simulations are set up without jets, but include the same cluster profile prescriptions and conduction rates for ease of comparison, while our experimental models include active AGN jets.

We ran this suite of simulations with varying input parameters to explore the effects of anisotropic conduction on the heat pump mechanism. In this paper we present results from a set of 11 of these simulations (along with two convergence test runs presented in \S\ref{sec:appendix}.) We assign acronyms to the simulations for ease of reference. C19 refers to the original \citet{Chen_2019} simulation with a jet and no conduction or magnetic fields (which was not re-analized here), and C19E is our equivalent comparison to the C19 model, with vanishingly small intracluster magnetic field strength (plasma beta of $\beta_{B} = 10^{6}$), but with the same refinement criteria and larger jet velocity of $v=0.2c$. The rest are as follows: jet, no conduction (JNC); jet, low conduction, $f_{\rm sp}=0.01$ (JLC); jet, high conduction, $f_{\rm sp}=0.1$ (JHC); jet with full anisotropic Spitzer-Braginskii conduction, $f_{\rm sp}=1$ (JSC); jet with isotropic conduction and $f_{\rm sp}=0.1$ (JHIC); no jet with isotropic conduction (NJIC); no jet with no conduction (NJNC); no jet with low conduction, $f_{\rm sp}=0.01$ (NJLC), no jet with high conduction, $f_{\rm sp}=0.1$ (NJHC); no jet with full anisotropic Spitzer-Braginskii conduction (NJSC); no jet with isotropic conduction and $f_{\rm sp}=0.1$ (NJHIC). A summary of their parameters, including these acronyms, is given in Table~\ref{tab:simulation-variables}. Due to the significant increase in computing time for the $f_{\rm sp}=1$ simulation by an order of magnitude compared to the simulations with $f_{\rm sp} \ll 1$, we only ran that simulation half the total time, until 520~Myrs, to sample at least one buoyancy oscillation cycle (see \S\ref{sec:dynamics}).

\section{Analysis \& Results}
\label{sec:results}

Consistent with previous AGN jet simulations in cluster environments \citep[e.g.][]{heinz2006,Morsony2010,Weinberger2017,Ehlert_2018,Chen_2019,Chen_2023,grete25}, our simulation shows jets that inflate buoyant cocoons and generate shock waves that propagate through the intracluster medium. As in \citet{Chen_2019}, we observe that the buoyant motions of the cavities and the circulation caused by the action of the jets drive the uplift of low entropy gas, which, in our simulations, reaches a maximum elevation of $\sim~100$~kpc. We find that the tangled magnetic field injected at startup drives some small-scale motions and induces anisotropies within the cluster temperature profiles as conduction along the field lines deviates from smooth radial temperature changes.

\subsection{Jet-Induced Buoyancy Oscillations}
\label{sec:dynamics}

\begin{figure*}[t]
  \centering
  \includegraphics[width=0.98\textwidth]{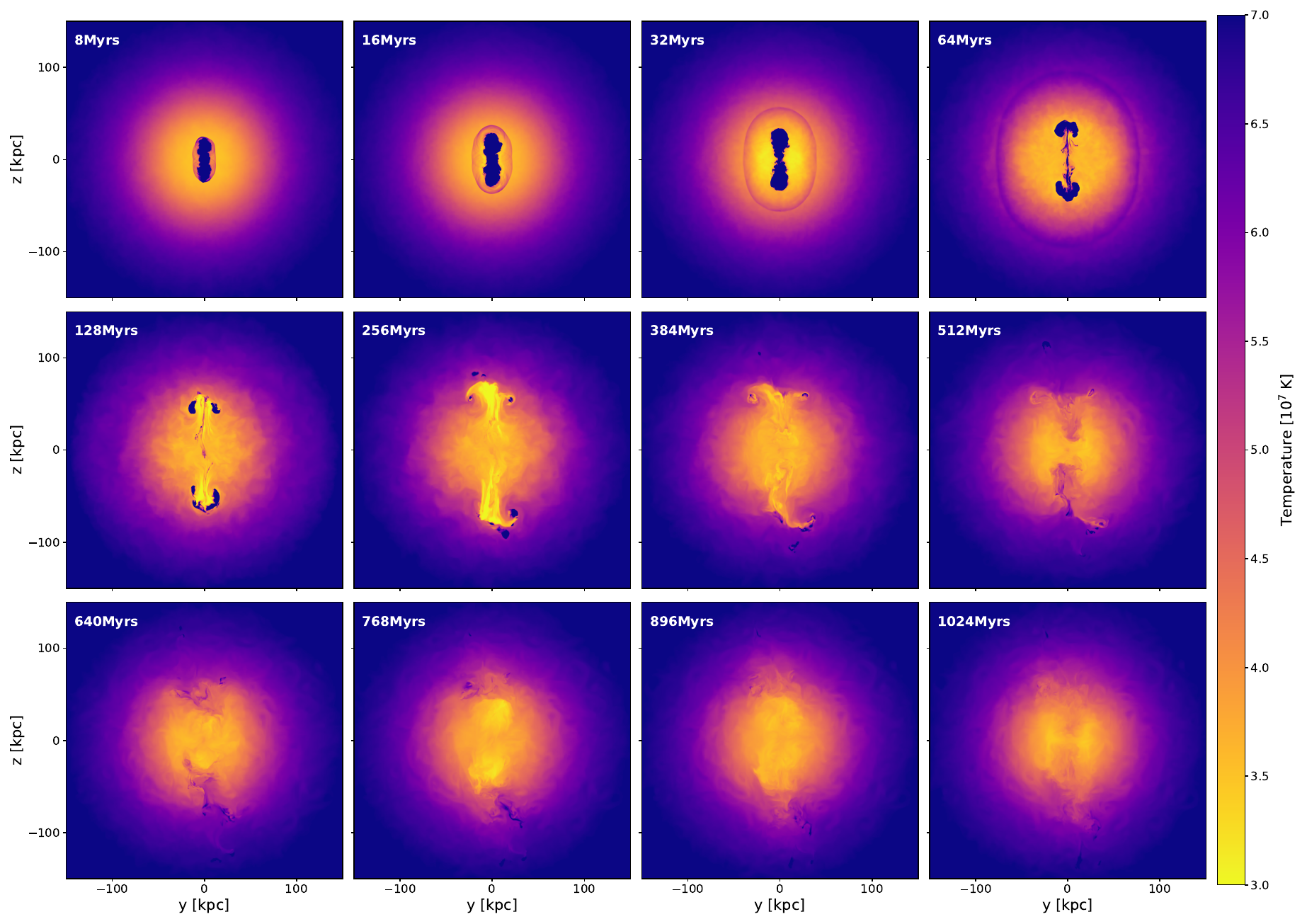}
  \caption{Temperature slices for the JHC run at the indicated times ranging from 8~Myrs to 1024~Myrs showing the evolution and uplift of cool core gas. The temperature color bar is scaled to highlight fluctuations in the thermal gas rather than the hot, non-thermal cocoon/cavity inflated by the jets.
  \label{fig:tempslices}}
\end{figure*}

By design, the dynamical evolution of the simulations follows the results described in \citet{Chen_2019} very closely.  We present snapshots throughout the JHC simulation, ranging from 8~Myrs to 1024~Myrs, in Fig.~\ref{fig:tempslices}. These are temperature slices through the center of the simulation box in the y-z plane (the mean jet axis is oriented along the z-axis.) After the jet turns off at 10~Myrs, the system is allowed to evolve over a total duration of 1024~Myrs. 

As in \citet{Chen_2019}, we observe a cyclic effect of uplift induced by the jet and the buoyant motions of the low density gas inside the cavities. The time slices at 128, 256, and 384 Myrs show clear uplift of low temperature gas. This cold, low-entropy gas is negatively buoyant and sinks back down towards the cluster center. The first buoyancy oscillation reaches bottom at approximately the 512 Myr time step, at which point the low entropy gas has sunk back down and, due to convective overshoot, adiabatic compression, and the effects of heat conduction (which has acted to increase the entropy of this gas), is therefore at higher temperature than the cool core, as outlined in the initial heat pump model \citep{Chen_2019}. We can see this inverted phase of the buoyancy cycle in the time slice for 512~Myrs. 

We then observe a smaller amplitude repetition of the uplift-downdraft cycle. Thus, the inflation of cavities and the subsequent uplift of cold gas excited buoyancy oscillations in the cluster. The characteristic frequency for these oscillations is the Brunt-V\"{a}is\"{a}l\"{a} frequency
\begin{equation}
     N = \sqrt{-\frac{g}{\rho}\frac{d\rho}{dr}} \sim \frac{1}{\tau_{\rm ff}}
\end{equation}
with a characteristic oscillation timescale of 
\begin{equation}
    \tau_{\rm osc} \sim \frac{2\pi}{N} \sim 2\pi \frac{R^{3/2}}{\sqrt{G M(<R)}} \sim 520 \,\text{Myrs}
    \label{eq:buoyancy}
\end{equation}
where $R~\sim~60~\text{kpc}$ is the approximate elevation of the uplifted gas at the peak of the first oscillation, $G$ is the gravitational constant, and $M(<R) \approx 1.4 \times 10^{44} \, \text{g}$ is the mass enclosed within $R$. By the end of the simulation at 1024~Myrs, the cluster has undergone roughly 2 oscillation cycles, as can be seen in Fig.~\ref{fig:tempslices} for the JHC simulation. 
 
\begin{figure}[h!t]
  \centering
  \includegraphics[width=0.47\textwidth]{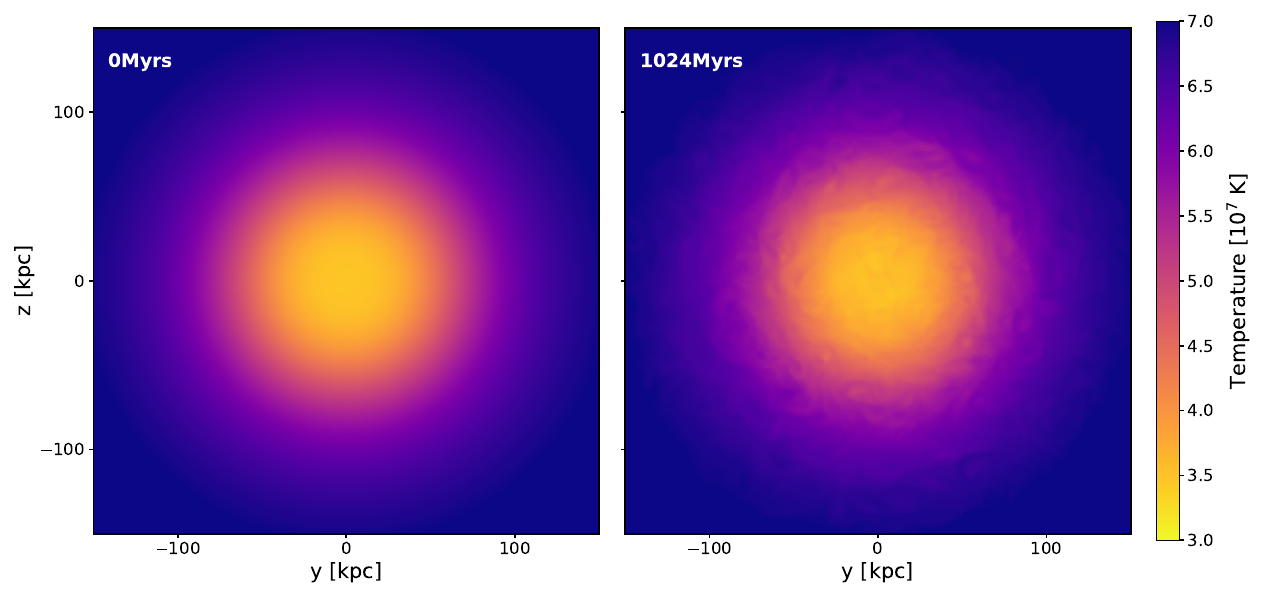}
  \caption{Temperature slices for the NJC run cluster core at 0 and 1024~Myrs. The small irregularities in the structure are due to magnetic field induced low level motions driving weak turbulence and conductive heat transport in random, non-radial directions, leading to small temperature anisotropies.}
  \label{fig:tempslices_nj}
\end{figure}

By construction, this effect is not apparent in the control simulations without jets, as there is no driven uplift of gas. The initial and final state of the NJHC are shown in Fig.~\ref{fig:tempslices_nj} and are dynamically undisturbed.

We note that the addition of magnetic fields to these cluster simulations drives small-scale irregularities that are more discernible in the NJHC control simulation, where large-scale dynamical effects driven by the jet are absent. This induces channels of heating which lead to anisotropies in the cluster's pressure and density profiles and resolution granularity in the temperature structure. These small scale, low amplitude effects can be observed in the right panel of Fig.~\ref{fig:tempslices_nj}. 

\begin{figure*}[h!t]
  \centering
  \includegraphics[width=0.98\textwidth]{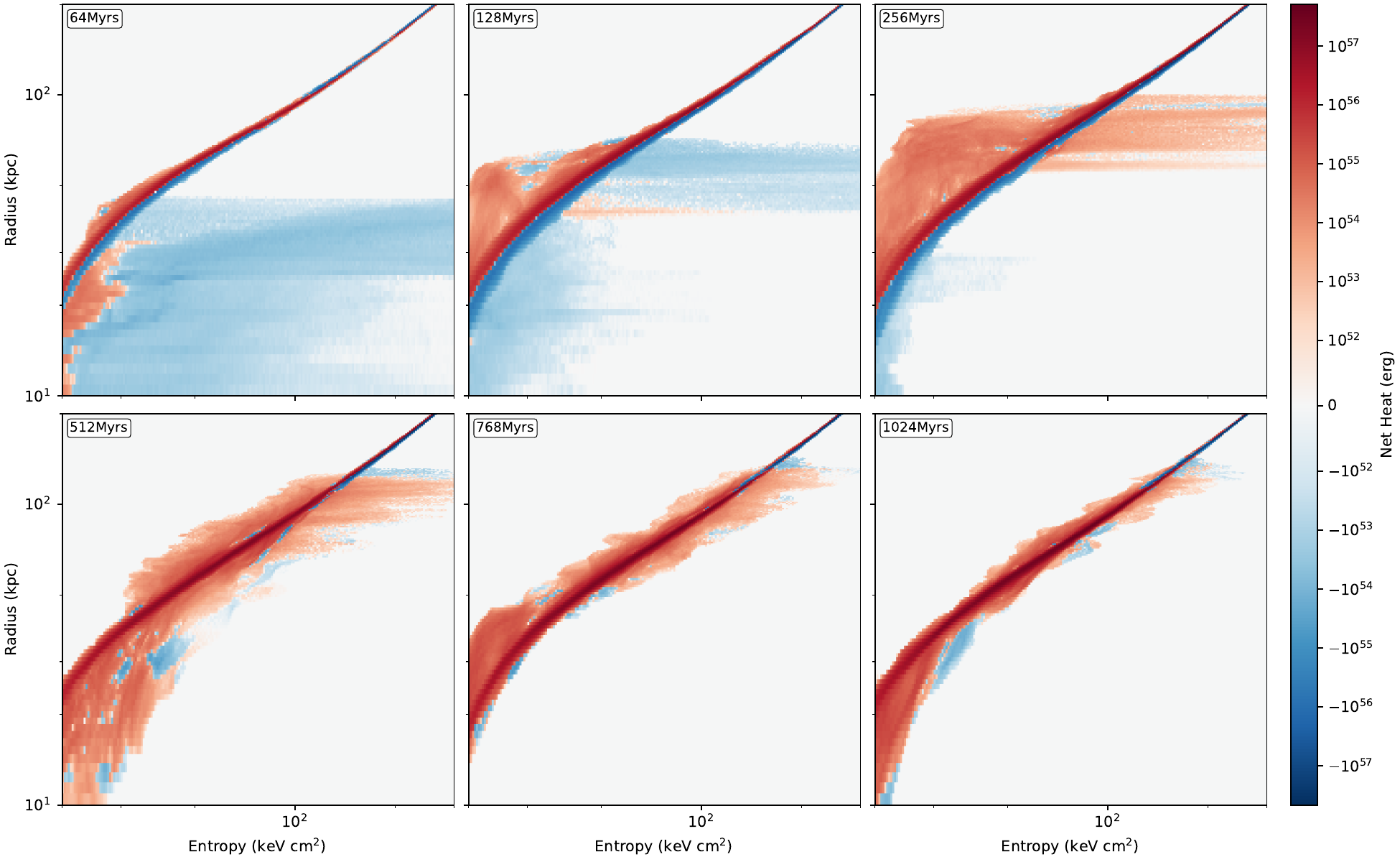}
  \caption{Phase plots (2D histograms) of net conductive heat $H$ as functions of radius and entropy for time steps at 64, 128, 256, 512, 768, and 1024~Myrs for the JHC simulation. The time sequence shows the cycling of lower entropy gas as it rises and falls throughout the simulation. Red colors indicated net conductive heating while blue colors indicate net cooling. The bright quasi-hyperbolic line corresponds to the mean cluster profile and can also be observed in Fig.~\ref{fig:heatphase_nj} for the control simulation. Gas above this line (in the upper left corner of the plot) corresponds to uplifted gas that has lower entropy than the average hydrostatic gas at the same radius. Gas below the line (lower right corner) corresponds to gas with excess entropy, either due to down-draft in the inward-moving part of the buoyancy cycle at 512~Myrs and 1024~Myrs, or shock heating (early frames). We are interested in how uplifted gas is heated by conduction, which we can quantify by summing over the (generally heated) gas above the equilibrium line, using the masks shown in Fig.\ref{fig:heatphase_nj}.}
  \label{fig:heatphase}
\end{figure*}

\begin{figure*}[h!t]
  \centering
  \includegraphics[width=0.98\textwidth]{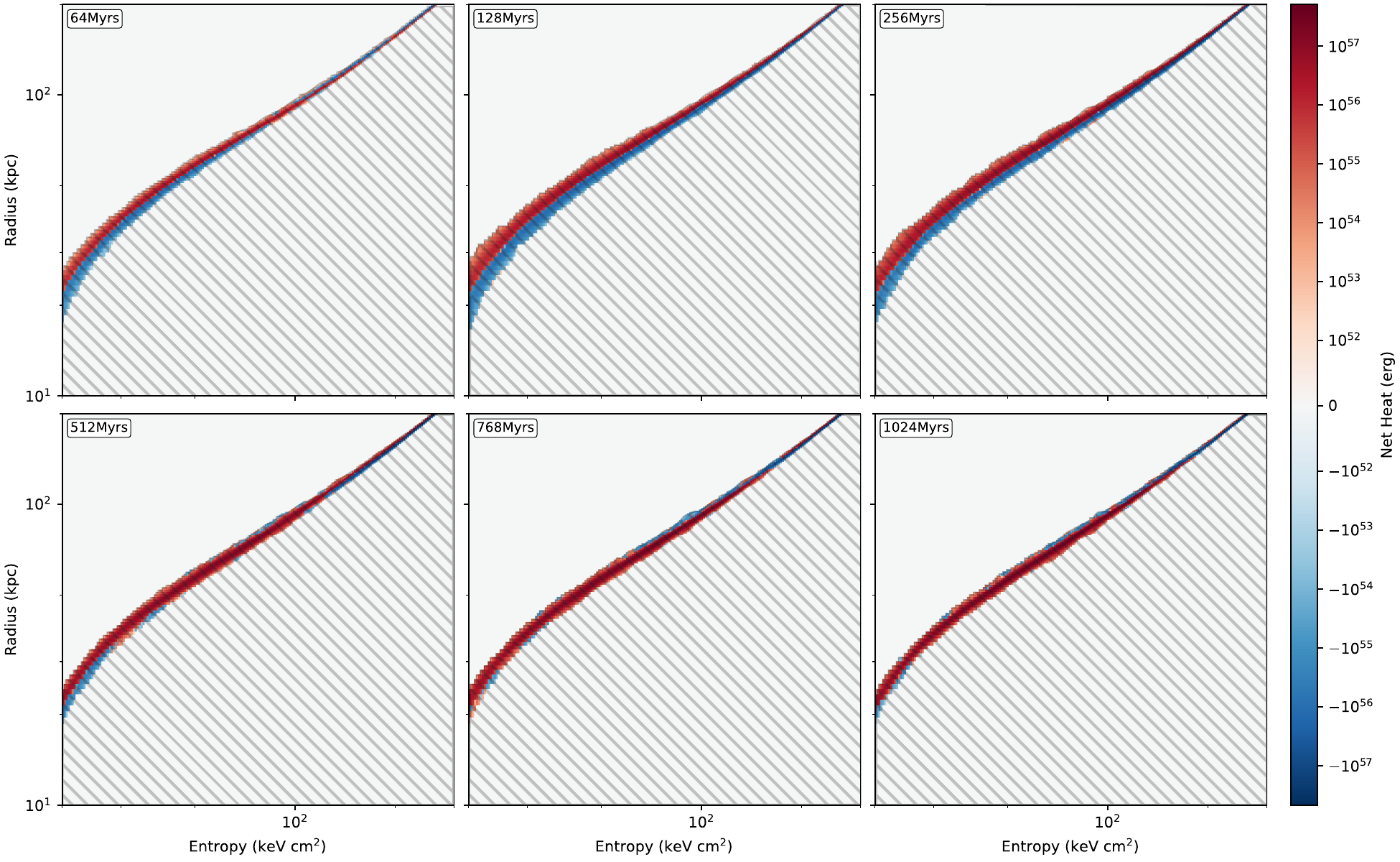}
  \caption{The same phase plot for $H$ as in Fig.~\ref{fig:heatphase} for the NJHC simulation. There is little change in the entropy profile over time without jet induced activity. The slight spreading at early times shows the initial smoothing of the temperature gradient through conduction along the randomized magnetic field channels. The hatched gray area shows the mask applied in the uplift analysis presented in \S\ref{sec:phase_analysis} which is used to isolate only uplifted low entropy gas (see Fig.~\ref{fig:heat_pump}).}
  \label{fig:heatphase_nj}
\end{figure*}

The jet-driven buoyancy oscillations (and the overall transport of gas) are further illustrated in Fig.~\ref{fig:heatphase}, which shows phase diagrams of the net heat $H$ deposited by conduction by radius and entropy bins for six different times. Red shading indicates net heating, blue shading indicates net cooling. The equivalent plots are shown for the NJHC simulation in Fig.~\ref{fig:heatphase_nj}. The implications for the thermal evolution of the cluster will be discussed further in \S\ref{sec:thermalevolution}. 

We see the same buoyant movement of the gas as demonstrated in Fig.~4 of \citet{Chen_2019}. The unperturbed cluster follows the bright central curves visible (below the hatched mask) in Fig.~\ref{fig:heatphase_nj}. Gas above this line represents uplifted low entropy gas and gas below the line indicates gas with excess entropy compared to the initial hydrostatic case, either due to buoyancy at late times or shock heating at early times. The buoyancy cycle is visible by the change in gas above and below the line with time.

\subsection{Thermal Evolution}
\label{sec:thermalevolution}

To assess the impact of the heat pump mechanism on the thermal evolution of the cluster core, we examine a range of diagnostics, including temperature and entropy profiles, heat deposition, and effects of magnetic fields on the system.

\subsubsection{Temperature Evolution of the Cluster}
\label{sec:temp}

As the main question addressed in this paper is to what extent the interplay of jet-induced uplift of cold gas and thermal conduction can heat cluster cores, we calculate radial entropy and temperature profiles over time for the simulations. 

\begin{figure}[h!t]
  \centering
  \includegraphics[width=0.47\textwidth]{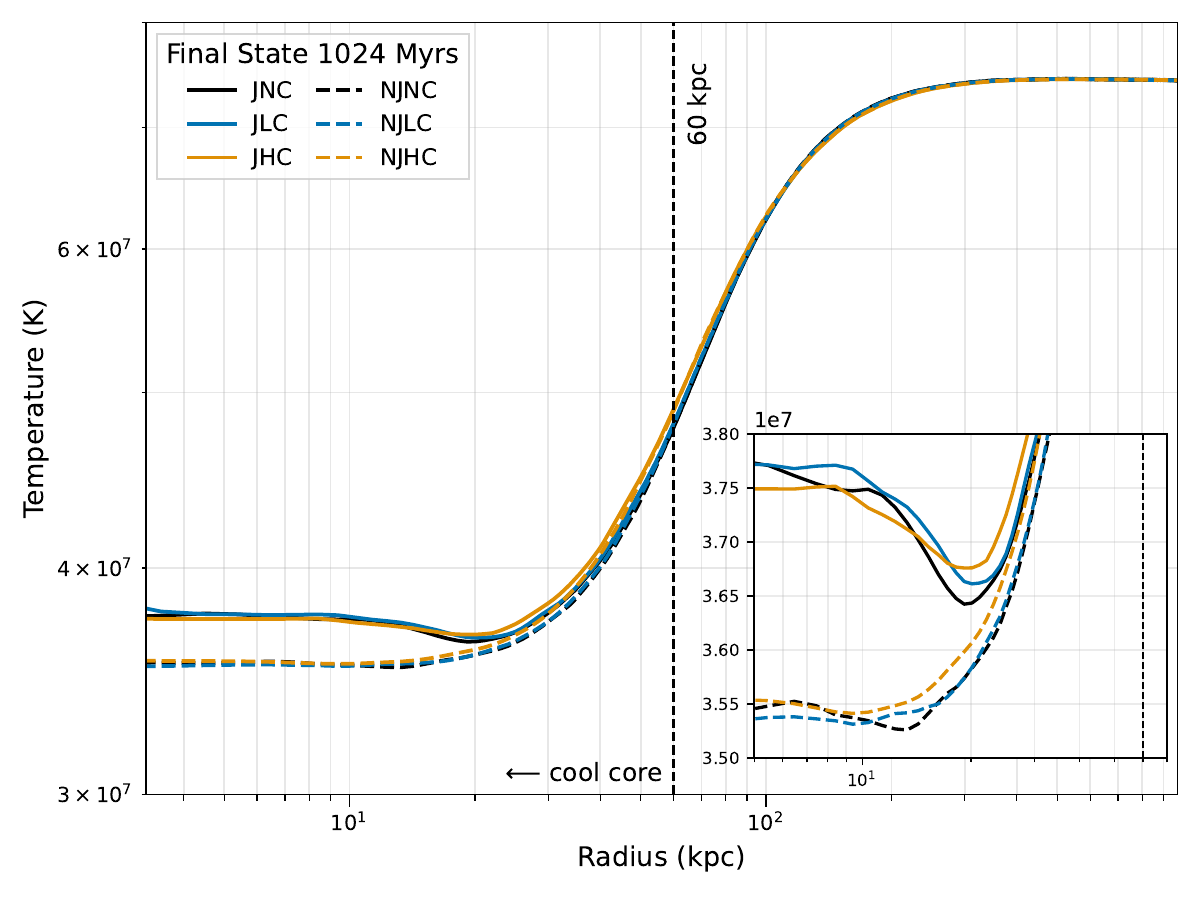}
  \caption{Radial temperature profiles for the 6 simulations at the final timestep, revealing a small amount of heating in the core for each of the simulations. The jet models (solid lines) show a measurable increase in core temperature, while the no jet models (dashed lines) show a more modest temperature increase. The inset plot is a zoomed in view  of the effect on the cluster core.}
  \label{fig:temp_final}
\end{figure}

Figure~\ref{fig:temp_final} shows a general increase in the central temperature in all cases. It is important to once again note that our simulations do not include radiative cooling, as our goal is to quantify the heat input by the heat pump mechanism. Similarly, it is important to note that even in the control simulation without conduction and without jet feedback, we see an increase in the temperature relative to the initial state due to the modest dissipation of some of the magnetic energy, as the random initial magnetic field relaxes into dynamical equilibrium and is affected by some numerical dissipation in the process. Since we are interested in the net effect generated by adding jet feedback and conduction, we will simply subtract this baseline effect and quantify the change relative to the background state (that is, our NJNC run serves as the baseline control simulation for all runs.)

\begin{figure}[h!t]
  \centering
  \includegraphics[width=0.47\textwidth]{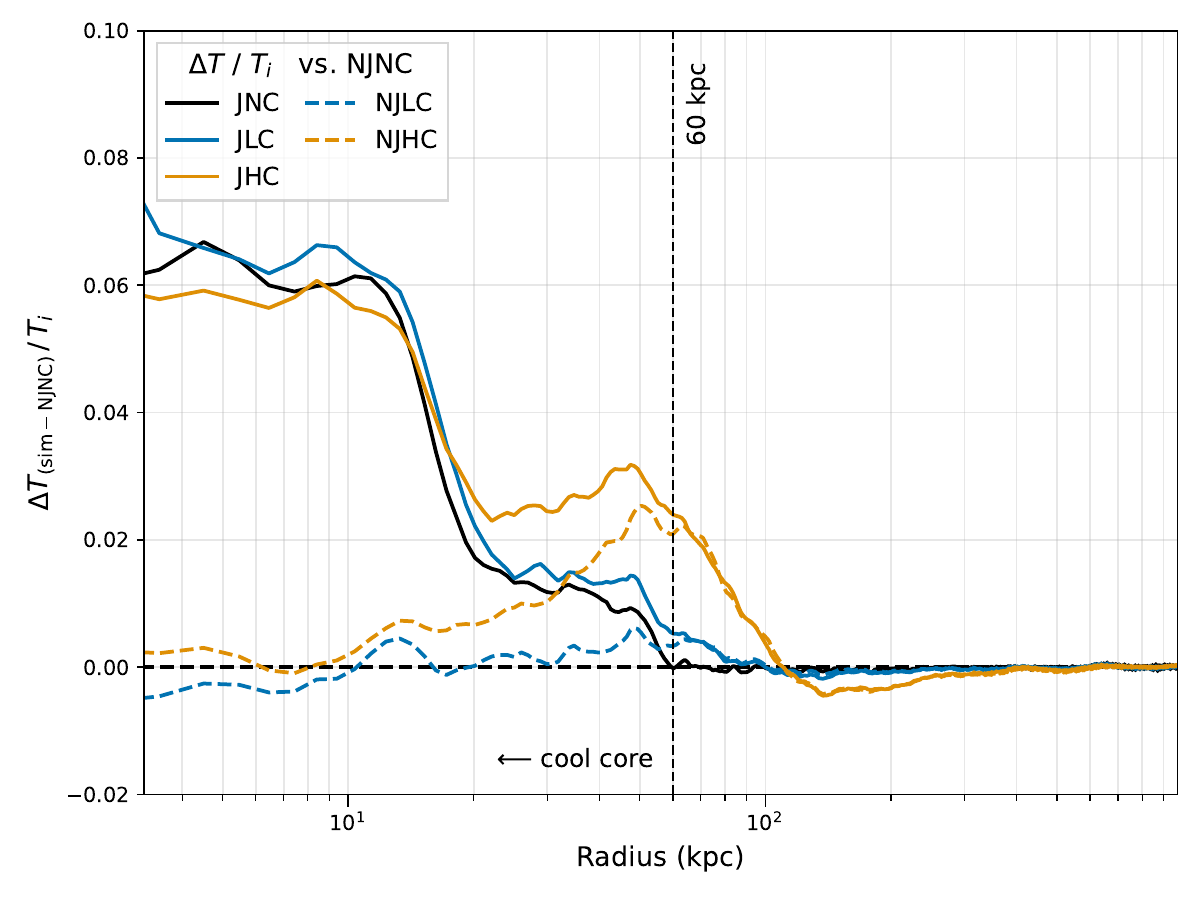}
  \caption{Fractional change in cluster temperature at the end of simulation, after 1024~Myr runtime, for 5 simulations, with the NJNC profile subtracted. We see an increase of order $\sim~6\%$ in cluster center for the three jetted simulations (solid lines in black, blue, and orange for the JNC, JLC, and JHC runs, respectively). The dashed lines are the corresponding control runs. Comparing runs with and without conduction, we see at most a $\sim~2\%$ gain from the heating generated by conduction in the temperature transition zone from 20~kpc to 100~kpc for the $f_{\rm sp}=0.1$ simulations (orange lines) and less than 1\% increase for the $f_{\rm sp}=0.01$ run. The net increase in temperature for the JHC simulation (solid orange) is comparable to the sum of the individual effects of the jet only, no conduction case (JNC; solid black) and the high conduction, no jet control (NJHC; dashed orange), indicating that the combined effect can be understood as a linear superposition of jet heating and conductive heating.}
  \label{fig:temp_ratio}
\end{figure}

As illustrated in Figure~\ref{fig:temp_ratio}, only a modest increase in temperature is seen in the JHC, JNC, and the NJHC simulations over the control. This is not surprising, given the 1\% duty cycle of the simulation and the suppression factor of 10 below the Spitzer-Braginskii value of our anisotropic conduction implementation. 

\subsubsection{Entropy Evolution of the Cluster}
\label{sec:entropy}

\begin{figure}[h!t]
  \centering
  \includegraphics[width=0.47\textwidth]{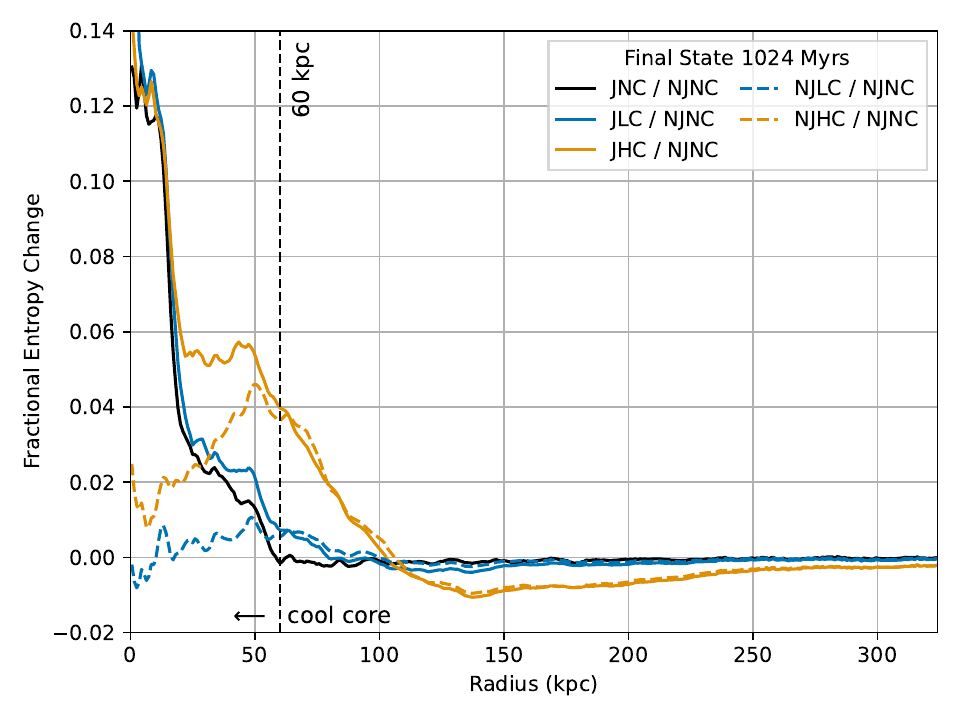}
  \caption{Fractional change of entropy as a function of radius $R$ after 1024~Myrs of evolution. The solid blue, orange, and black curves show the JLC, JHC, and JNC runs and the dashed blue and orange lines show the NJLC and NJHC runs, respectively. All curves show the fractional excess relative to the NJNC run. Within the cool core region, conduction accounts for about $4\%$ of the entropy change in the temperature transition region between 20~kpc and 100~kpc, while the jet accounts for $12\%$, concentrated mostly in the innermost 20~kpc.
  As in Fig. 7, we find that the effect of the JHC simulation (solid orange) can be understood as the sum of the individual contributions from the jet only, no conduction case (JNC; solid black) and the high conduction, no jet control (NJHC; dashed orange).}
  \label{fig:frac_entropy}
\end{figure}

Figure~\ref{fig:frac_entropy} shows the breakdown of the individual effects on entropy. We see the fractional ratio of the change in entropy with radius for three simulations, JHC, JNC, and NJHC, which are all normalized by the initial phase of the NJNC simulation. The blue curve (JHC) shows the combined effect on entropy by the jet and conduction, while the green and orange curves show the individual jet (JNC) and conduction effects (NJHC), respectively.  Because our JSC simulation only ran to 512~Myrs due to computational limits, we do not include results from the JSC and NJSC simulations in this figure.

The figure shows increased entropy in the core for all three simulations, of order 5-10\%, compared to the NJNC one. Generally, the entropy increase is largest for the simulation that includes both jet feedback and thermal conduction. The JNC simulation shows roughly the same increase in the very center of the simulation, but falls below the JHC simulation outside of about 20~kpc. The NJHC simulations show an increase in entropy very similar to the JHC simulation outside of the cool core.  

The overall effect of including both jets and thermal conduction is to increase entropy relative to simulations that include only one or the other effect. Not surprisingly, simulations with both jet feedback and heat conduction lead to the largest increase in entropy in the cool core. However, Fig.~\ref{fig:temp_final} also shows that there appears to be no {\em additional} increase {\em beyond} the sum of both effects individually, contrary to what might be expected in the presence of a strong heat pump effect.

Importantly, our simulations also support the initial conclusion that thermal conduction is inefficient at heating the very centers of cool core clusters. This inefficiency is due to the reduced conduction rates at lower temperatures and the absence of a strong temperature gradient, which was one of the main reasons that thermal conduction has been dismissed as a feedback mechanism \citep[e.g.][]{Zakamska_2003}. The changes in entropy and temperature are mostly limited to the transition region between $50\,{\rm kpc}$ and $100\,{\rm kpc}$, where the temperature is larger and the temperature gradient shows strong curvature.

\subsection{Heat Conduction Efficiency}
\label{sec:gain}

Our simulations cover a single episode of jet activity, following the same duration and power as in \citet{Chen_2019}, as we wish to test the model put forward in that paper. The total energy injected into the grid by the jet is thus
\begin{equation}
E_{\rm inj} = 10^{45} \, \mathrm{erg\,s^{-1}} \times 10\,\mathrm{Myr}
  \approx 3.15\times 10^{59} \,\mathrm{erg}
\end{equation}

\begin{figure}[h]
  \centering
  \includegraphics[width=\columnwidth]{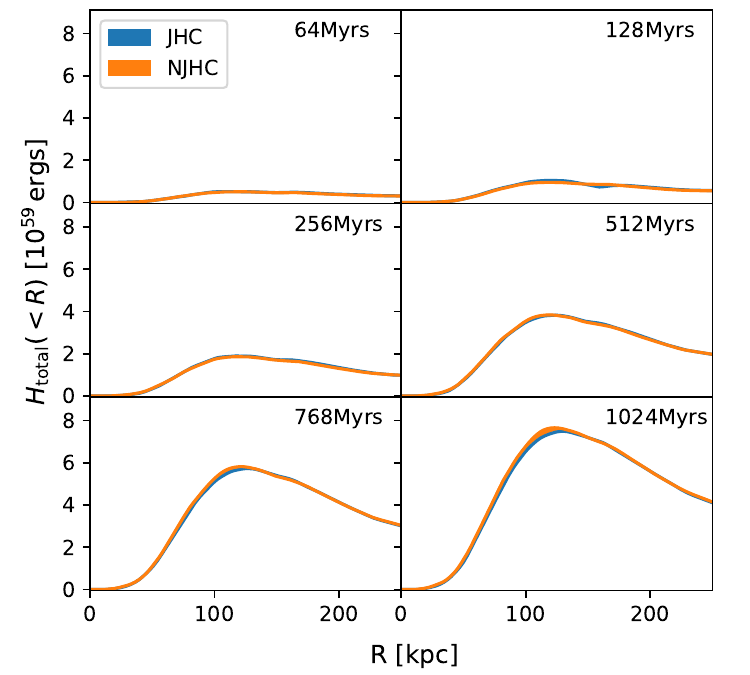}
  \caption{Net total heat deposited within a sphere of radius $R$ as a function of $R$ for the same output times used in Fig.~\ref{fig:heatphase}. The JHC simulation is shown in blue, while the NJHC model injection is shown in orange. The peak at around 120~kpc indicates net heating inside this radius and net cooling outside.}
  \label{fig:total_heat}
\end{figure}

We can compare this to the net heat transported in the cool core throughout the simulation. We plot the total heat deposited by conduction $H(<R)$ into a sphere of radius $R$ in Fig.~\ref{fig:total_heat}. In both simulations, this function increases with time and has a peak at approximately 120 kpc radius, outside of which it decreases. This indicates that inside this radius, conduction {\rm increases} the temperature, and outside of it, conduction {\rm decreases} the temperature, as expected in a thermally stratified atmosphere. 

After a gigayear, we find that, in both JHC and NJHC runs, the total amount of heat energy deposited is of order $H_{\rm core} \sim 8\times 10^{59}\,{\rm ergs}$, for an average heating rate of
\begin{equation}
    P_{\rm cond,0.1sp}\sim 2.5\times 10^{43}\,{\rm ergs/s}
\end{equation}
This is comparable to the energy injected by a single, moderate episode of AGN activity as simulated here. 

An increase in the conduction coefficient from the assumed value of $f_{\rm sp}=0.1$ in these two runs would increase this energy injection accordingly. At most, we might therefore expect a tenfold increase in heating, if anisotropic conduction operated at the full Spitzer-Braginskii rate.

\begin{figure}[t]
  \centering
  \includegraphics[width=0.47\textwidth]{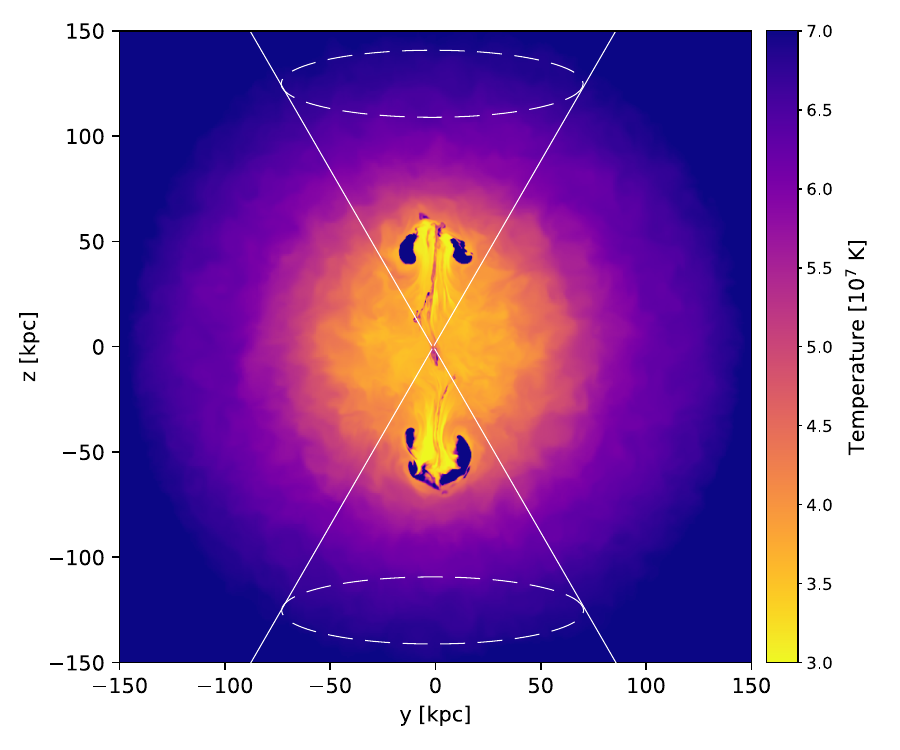}
  \caption{Schematic overlay of our $30^{\circ}$ analysis cone on a temperature slice of the JHC run at 128~Myrs. We use both $30^{\circ}$ cones and a $15^{\circ}$ cones in subsequent analyses and corresponding anti-cone ``ruff'' regions at varying angles, as indicated in the respective figures.}
  \label{fig:cone}
\end{figure}

\section{Discussion}
\label{sec:discussion}

\subsection{Quantifying the Heat Pump Efficiency}

Comparison of the JHC and NJHC simulations in Fig.~\ref{fig:total_heat} shows that, under conservative assumptions, no net heat excess inside the cluster core is measurable in our simulations. This, and the similarity of the net increase of the temperature and entropy profiles due to conduction (i.e., the similarity of JHC curves and the sum of the NJHC and JNC increases in Figs.~\ref{fig:temp_final}, ~\ref{fig:temp_ratio}, and \ref{fig:frac_entropy}) indicates that no strong heat pump effect is apparent in the fiducial JHC simulation vis-a-vis the control simulations. In other words, the net conductive transport of heat into the cluster core is not substantially increased by the action of jets in the numerical experiments conducted here. Thus, our simulations do not support the hypothesis that large gains in feedback efficiency can be achieved by the heat pump mechanism outlined in \citet{Chen_2019} under the same experimental conditions of the numerical experiment presented therein. That is, adding anisotropic thermal conduction at 1\% or 10\% of the Spitzer-Braginskii rate does not lead to substantial additional heating by the heat pump mechanism.

Given the limitations of the simulations, it is plausible that more favorable experimental conditions---most notably a larger lift distance of the low entropy gas by more powerful jets or a longer injection episode---may lead to a more favorable outcome, and we leave investigation of this possibility to future work.

Before we address the factors we identify as suppressing the efficiency of the heat pump mechanism, we aim to quantify how much heat is, in fact, deposited into the uplifted gas. Because lateral conduction, i.e., heat conduction perpendicular to the radius vector in the cluster, does not change the radial distribution of heat---and will therefore not appear as a large difference in radial heat profile shown in Fig.~\ref{fig:total_heat}---we need to isolate the conduction of heat into low entropy gas. We perform two analyses to this end.

\subsubsection{Cone Sections}
As jets mostly drive transport along the jet axis (the z-axis in our specific setup), we select conical regions in the positive and negative z-directions out of a sphere of radius $R$ to enclose the area of interest. The vertices of the cones are positioned at the center of our box with opening angles of $30^{\circ}$ and $15^{\circ}$ around the jet axis. We then perform diagnostic tests of variables related to the heat pump mechanism inside these cones. This includes as the total heat deposition $H$ by conduction over time in this conical region for JHC, compared to the same diagnostics for the NJHC simulation. We also compare the cone diagnostic to the equivalent inverse spherical selection (i.e.,  sphere with a corresponding conical void), further referred to as the ``ruff'' region. The cone region is shown as an overlaid schematic in Fig.~\ref{fig:cone}.

\begin{figure}[t]
  \centering
  \includegraphics[width=\columnwidth]{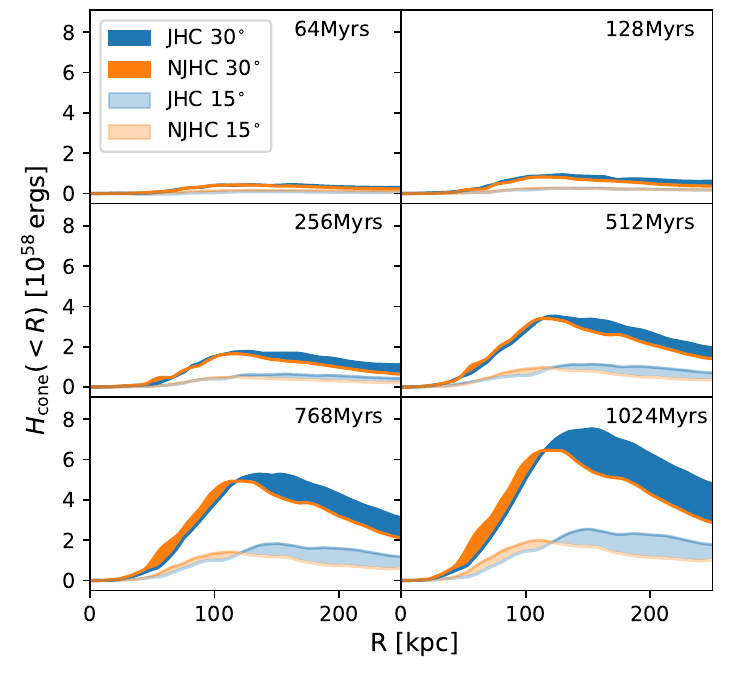}
  \caption{Net heat deposited within a sphere of radius $R$ as a function of $R$ inside a 30 degree  (dark colors) or 15 degree (light color) cone around the mean jet axis for the same output times used in Fig.~\ref{fig:heatphase}. The JHC simulation is shown in blue, while the NJHC model injection is shown in orange. The jet simulation shows a clear excess of conductive heating within the jet cone at the transition from cool core to hot atmosphere (at around 100 kpc) while the very center of the jet cone shows a lower amount of conductive heating.}
  \label{fig:jet_heat}
\end{figure}

To assess the heat pump efficiency, we calculate the excess heat deposited by conduction as a function of radius $H(<R)$ inside this cone, which we plot in Fig.~\ref{fig:jet_heat} for both the JHC and NJHC runs in blue and orange, respectively. The shaded areas indicate the excess of one simulation over the other. Unsurprisingly, the NJHC simulation exhibits the same functional form in both Fig.~\ref{fig:total_heat} and Fig.~\ref{fig:jet_heat}, with the appropriate volume scaling factors, given that it is roughly spherically symmetric. 

However, the JHC simulation shows a clear excess, indicating that additional heat is deposited into the jet cone at radii between $120 {\rm~kpc}$ and $150\,{\rm~kpc}$. Overall, the JHC simulations shows an excess of 
\begin{equation}
    \Delta H_{\rm 30^{\circ}} \sim~2 \times 10^{58}\,{\rm ergs}
    \label{eq:heat_pump}
\end{equation}
deposited into the jet cone. Since the total integral of $\int dV H$ over the simulation volume must vanish, this implies a net heat transfer from outside of this cone into the uplifted low entropy gas, as expected from the heat pump model.

\subsubsection{Entropy-Radius Phase Analysis}
\label{sec:phase_analysis}
\begin{figure}[t]
  \centering
  \includegraphics[width=\columnwidth]{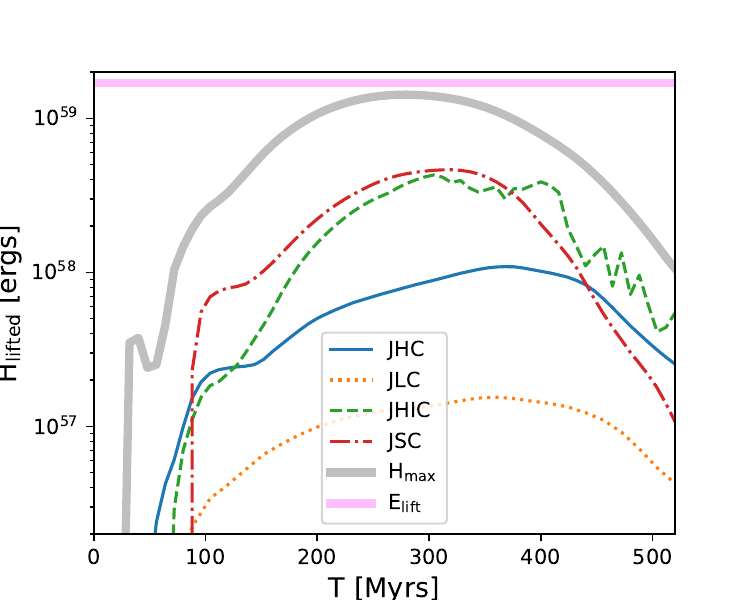}
  \caption{Net heat $H$ deposited into uplifted low entropy gas for the JHC (blue), JLC (orange), JSC (dash-dotted red), and JHIC (dashed green) simulations. The quantity is calculated summing over the net heat deposition shown in the 2D histogram (phase plot) in Fig.~\ref{fig:heatphase} after applying the masks shown in Fig.~\ref{fig:heatphase_nj}. The figure demonstrates the effect of the buoyancy oscillations, as material initially uplifted and conductively heated then falls below its hydrostatic equilibrium radius, thus, all curves peak at around half of the oscillation time $\tau_{\rm osc}$ from eq.~\ref{eq:buoyancy}. This implies that these curves for $H$ are lower limits, as not all initially low entropy material will be included in the sum. The maximum possible amount of heat $H_{\rm max}$ transferred by the heat pump (given the amount of uplift generated by the simulated jets using eq.~(15) from \citet{Chen_2019} is overplotted as a thick gray line. The total gravitational lift energy expended by the jets $E_{\rm grav}$ (see Fig.~\ref{fig:grav}) is shown as the solid magenta line.}
  \label{fig:heat_pump}
\end{figure}

To complement this analysis, we also calculate the net heat deposited into the uplifted low entropy gas from the phase diagram in Fig.~\ref{fig:heatphase}. To this end, we construct masks that isolate only gas that, at a given cluster radius, has lower entropy than any gas in the equivalent control simulation. The masks are shown as hatched areas in Fig.~\ref{fig:heatphase_nj}. We then calculate the total net heat $H_{\rm lifted}$ in the unmasked area an plot it in Fig.~\ref{fig:heat_pump}. The figure shows there is net excess heat deposited into the uplifted gas, once again indicating the action of the heat pump. 

Because of the buoyancy oscillations induced in the cluster, the total value fluctuates between the two peaks. The downward phase of the oscillation moves the gas below the equilibrium line in the phase plot, as can be seen in Fig.~\ref{fig:heatphase} for the 512~Myr and 1024~Myr panels. Thus, we consider this measure of heat pump energy transfer a lower limit. Because the buoyancy oscillations damp out over time, this signal becomes weaker during the second oscillation cycle. We thus limit this figure to simulation times below 512~Myrs. Therefore, we can include the JSC and the JHIC simulations in this analysis.

For our fiducial JHC simulation, the peak of the figure reaches a value of 
\begin{equation}
    \Delta H_{\rm lift,JHC} \sim 1.25\times 10^{58}\,{\rm ergs}
\end{equation}
comparable to the value independently determined in eq.~(\ref{eq:heat_pump}).  

For the more optimistic JSP and JHIC simulations, we find an increase by roughly a factor of five over the JHC simulation, with
\begin{equation}
    \Delta H_{\rm lift,JSC} \sim  \Delta H_{\rm lift,JHIC} \sim 1.25\times 10^{58}\,{\rm ergs}
\end{equation}
This is less than the direct ratio expected if the total heat were to scale linearly with $f_{\rm sp}$, since thermal conduction is generally self-limiting once conduction is sufficiently strong to change the temperature distribution materially.

Conversely, the JLC simulation shows a factor of ten decrease compared to the JHC simulation, in line with the expectation that, at small $f_{\rm sp}$, the heat transfer should scale linearly with $f_{\rm sp}$, as long as changes in temperature are small.

\subsubsection{Efficiency Estimates}
\label{sec:efficiency}
For Figs.~\ref{fig:jet_heat} and \ref{fig:heat_pump}, we conclude that, for our fiducial run with a value of $f_{\rm sp}=0.1$ and a 10~Myr jet injection at $10^{45}\,{\rm ergs/s}$, the long term (hundreds of Myrs) average heat transfer into uplifted low entropy gas is of order $1-2\times 10^{58}\,~\text{ergs}$. Since the total energy injected by the jet is $3.15 \times 10^{59}\,{\rm ergs}$, the relative energy gain due to the heat pump is only of order $\eta \sim 3-7\%$. This heat is extracted from laterally adjacent higher entropy gas, so the overall net effect on the average cluster temperature and entropy profiles is very small, as noted in \S\ref{sec:thermalevolution}. Larger and smaller values of $f_{\rm sp}$ naturally lead to larger and smaller values in $H_{\rm lifted}$, respectively.

To evaluate the efficiency of the heat pump mechanism, we compare this to the {\em maximum possible} energy {\em gain} $H_{\rm max}$, which is achieved if all the uplifted gas is thermalized to the equilibrium temperature at its radius, as discussed in \citet{Chen_2019}.

We evaluate this using the same masked selection of uplifted gas shown in Fig.~\ref{fig:heatphase_nj} and calculate the energy differential between the gas temperature differential and the equilibrium temperature $T_{\rm eq}$ from eq.~\ref{eq:temperature} [and following the formalism in eq.~(15) in \citet{Chen_2019}]: 
\begin{equation}
    H_{\rm max} = \int dV \frac{1}{\gamma-1}nk\left(T-T_{\rm eq}\right)
\end{equation}

The result is shown as the thick gray line in Fig.~\ref{fig:heat_pump}. This line peaks around 
$H_{\rm max} \sim 1.4\times 10^{59}\,{\rm ergs}$, and we take this to be the maximum achievable heat pump energy {\em gain} for the simulation conditions. Note that this value will depend on jet power and duty cycle. We thus find that our fiducial run reaches an efficiency between 10\% and 20\% of this value.

\begin{figure}[t]
  \centering
  \includegraphics[width=\columnwidth]{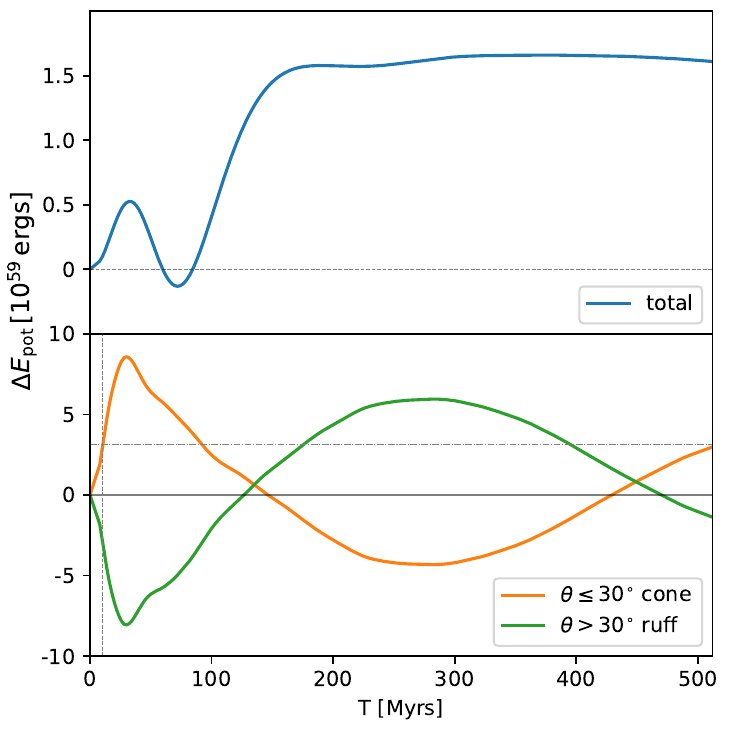}
  \caption{{\em Top:} Net excess gravitational potential energy as a function of time in the simulation with a jet compared to the unperturbed cluster in blue in units of $10^{59}\,{\rm ergs}$; {\em bottom:} net excess gravitational potential energy in the jet cone ($\theta\leq 30^{\circ}$) in orange and in the ``ruff'' ($\theta > 30^{\circ}$ in green. The cluster acts like an elevator counterweight, with the uplift along the jet axis being almost perfectly canceled by the downdraft in the perpendicular direction.}
  \label{fig:grav}
\end{figure}

Formally, the net {\em efficiency} of a heat pump can be quantified as the fraction of heat gained from the heat bath over the mechanical energy expended to lift the gas against the gravitational force of the cluster. Figure~\ref{fig:grav} plots the net excess gravitational potential energy as a function of time in the simulation. We find the amount of energy going into uplift is $E_{\rm grav}\approx~1.5 \times 10^{59}\,$~ergs, or about half of the total energy injected by the jet $E_{\rm tot}$ over the simulation duration.  The {\em maximum} possible heat pump efficiency {\em for our particular simulation setup} (jet power and duty cycle) is thus 
\begin{equation}
    \eta_{\rm max} \equiv H_{\rm max}/E_{\rm grav} \sim 82\%
\end{equation}
This value will change with different average lift elevation or different cluster temperature profiles.

Given our estimate of $H_{\rm JHC}$ for our fiducial JHC run from above, we thus find a long-term heat pump efficiency of $\eta_{\rm JHC}=H_{\rm JHC}/E_{\rm grav} \sim~6\%-12\%$. This value represents the efficiency boost in the system compared to the energy injected by the jet itself. 

For the smaller value of $f_{\rm sp}$ applied to the JLC simulation, the net gain is correspondingly smaller, as shown by the orange curve in Fig.~\ref{fig:heat_pump} for the JLC simulation. In this case, the efficiency is reduced by about an order of magnitude to about 1\% of the maximum possible rate, as expected from the reduced conduction rates. At even lower values of $f_{\rm sp}$, we expect the dependence to be linear. We thus conclude that, for $f_{\rm sp}\ll 1$, we can crudely approximate the heat pump efficiency as
\begin{equation}
    \eta_{\rm f_{\rm sp}\ll 1} \sim f_{\rm sp}\eta_{\rm max}
\end{equation}

Conversely, larger $f_{\rm sp}$ yield greater gains: The optimistic case of anisotropic conduction at the Spitzer-Braginskii rate (JSC) reaches a net efficiency of approximately 
\begin{equation}
    \eta_{f_{\rm sp}=1} \sim 40\% \sim 0.5 \eta_{\rm max}
\end{equation}
of the maximum possible value, given the net heat deposition of $H_{\rm JSC} \approx 5\times 10^{58}\,{\rm ergs}$.

Similarly, we find that isotropic conduction at $f_{\rm sp}$ in our JHIC run leads to larger heat pump efficiencies of $f_{\rm sp}=0.1$ with efficiencies of $\eta_{\rm JHIC} \sim 50\%$ of the maximum possible value, similar to the JSC case.

We can reasonably expect that longer or more powerful jet episodes would lead to increased lift with correspondingly large heat transfer efficiencies and increases in $\eta_{\rm max}$. However, as the simulations presented here were designed to test the initial conjecture of \citet{Chen_2019} at face value, we kept the overall simulation parameters consistent with that study, as outlined in \S\ref{sec:methods} and leave exploration of wider ranging parameter assumptions to future studies.

In conclusion, we find that the heat pump model as proposed in \citet{Chen_2019} requires optimistic assumptions about the level of suppression of conduction below the Spitzer-Braginskii value or the isotropy of conduction to yield meaningful increases in cluster heating. Such conditions might arise if turbulent conduction is efficient and/or kinetic suppression factors of the conductivity along the field are less severe than suggested in \citet{Roberg-Clark_2016, Roberg_Clark_2018}.

\subsection{Magnetic Draping Effects on Anisotropic Conduction}
\label{sec:draping}

Thermal conduction suffers from two major drawbacks as a heating agent in cool core clusters: The strong temperature dependence, which reduces heating of lower temperature gas---thus, making it inefficient at heating the gas most in need of heating---and the lack of conduction across magnetic field lines. The former can be overcome by longer heating time scales, which is why we extended our simulations to run for as long as 1 Gyr, but the latter is a limitation that is exacerbated by time. This is because temperature gradients in anisotropic conduction will naturally misalign over time.

In the case of the heat pump mechanism, an additional effect appears to limit conduction into the uplifted low entropy gas, which we will explore below. The dynamical stretching of magnetic field lines along the direction of uplift---commonly referred to as magnetic draping \citep[e.g.][]{Ruszkowski2007,Dursi2008}---causes the field lines, which are roughly radial in this region, to become misaligned with the primarily tangential temperature gradient induced by the uplift. The deleterious impact of magnetic draping on transport processes in galaxy clusters is a well studied phenomenon \citep[e.g.][]{Ettori2000,Vihlinin2001,Asai2004,Lyutikov2006,Ruszkowski2007,Ruszkowski2008}. We also note that \citet{Karen_Yang_2016} found similar impacts on conductive heat transport due to magnetic field geometries. Below, we will quantify how magnetic draping by jet-induced uplift in our simulations affects thermal conduction.

In order to illustrate the evolution of the ICM's magnetic field in response to the jet activity, we plot 2D projections of the magnetic field lines in Fig.~\ref{fig:bfieldslices}, overlaid on the temperature slices from Fig.~\ref{fig:tempslices} at four different times. This figure showcases the initial randomized field topology with a spectrum of characteristic lengths scales from about 10 kpc to 100 kpc, as per the chosen power spectrum, and the evolution can be tracked over the duration of the simulation. 

We find that, at later times, the magnetic field lines in the jet simulations are stretched and draped around the uplifted gas as it propagates through the cluster core, preferentially in the radial direction. As shown in Fig.~\ref{fig:bfieldslices}, the field lines tend to align perpendicular to the temperature gradient around the cooler gas.

\begin{figure}[t]
  \centering
  \includegraphics[width=0.98\columnwidth]{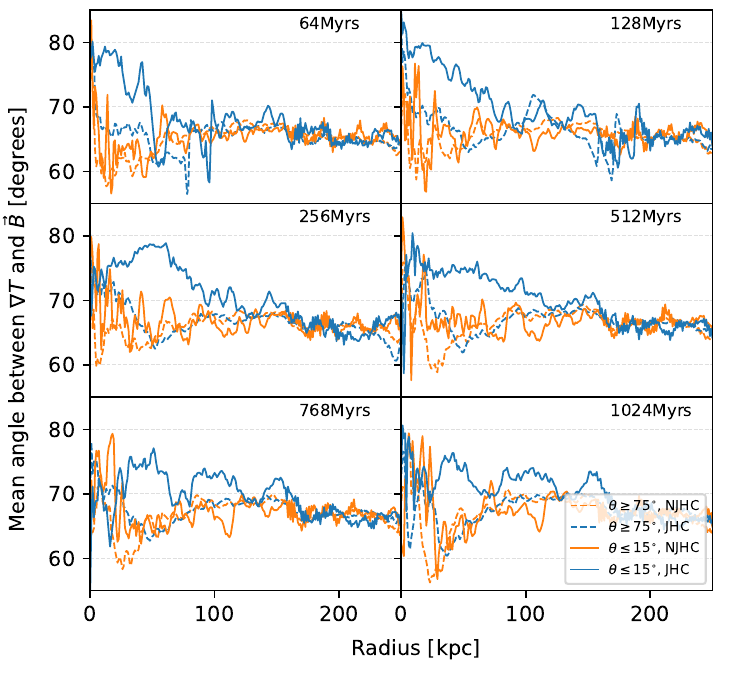}
  \caption{Misalignment between temperature gradient and B-field vector. Shown is the mean angle between $\nabla{T}$ and $\vec{B}$ in a $\theta \leq 15^{\circ}$ cone around the mean jet axis (solid lines) and in the ``ruff'' region $\theta \geq 75^{\circ}$ (dashed lines) for the JHC simulation (blue), compared to the NJHC simulation (orange) at different times. The figure shows a clear increase in the misalignment between $\nabla{T}$ and $\vec{B}$ in the JHC simulation within the cool core ($r\lesssim 150\,{\rm kpc}$) in the jet cone, while no such large difference in the misalignment between the JHC and the NJHC simulation is visible in the ``ruff'' region. The figure shows a secular increase in the misalignment angle with time, due to the fact that anisotropic conduction acts to misalign the temperature gradient with the local B-field direction.}
  \label{fig:misalignment}
\end{figure}

We can quantify this effect further: In Fig.~\ref{fig:misalignment} we show the mean angle between the temperature gradient and the B-field vectors for the JHC and NJHC simulations, within the $15^{\circ}$ jet cone and the equatorial ``ruff'' of $\theta \geq 75^{\circ}$. Within the cone region (solid lines), we see that the misalignment of the JHC simulation grows over time, as the mean angle between the vectors increases (solid blue lines). A larger angle implies that the two vectors are becoming more perpendicular/less aligned. We see that, for the NJHC run (orange lines), the angles in the jet and the ``ruff'' regions (solid and dashed lines, respectively) do not exhibit the same trend. The same is true for the ``ruff'' region of the JHC simulation. This implies that the jet is driving this self-limiting behavior, turning the vectors perpendicular, and thus blocking conduction along the temperature gradient. Interestingly, all simulations show mean angles larger than the random average angle of $60^{\circ}$, indicating that anisotropic conduction does indeed increase the misalignment between B-field and temperature gradient.

\begin{figure}[t]
  \centering
  \includegraphics[width=0.98\columnwidth]{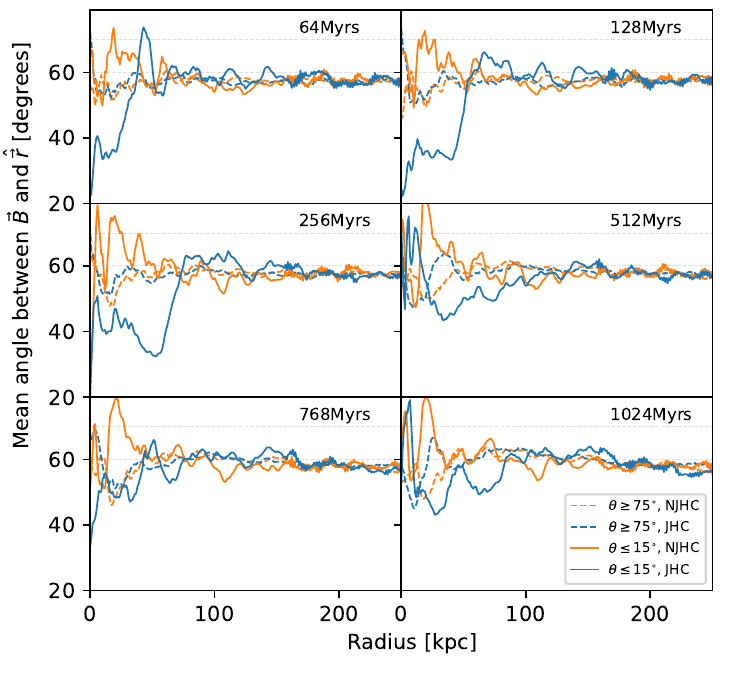}
  \caption{Mean angle between B-field vector and the radius vector, measuring whether the field is predominantly radial or tangential, in a $\theta \leq 15^{\circ}$ cone around the mean jet axis (solid lines) and in the ``ruff'' region $\theta \geq 75^{\circ}$ (dashed lines) for the JHC simulation (blue), compared to the NJHC simulation (orange) at different times. The JHC simulation shows a more radial field alignment within the jet cone compared to the NJHC simulation.}
  \label{fig:b_angle}
\end{figure}

\begin{figure}[t]
  \centering
  \includegraphics[width=0.98\columnwidth]{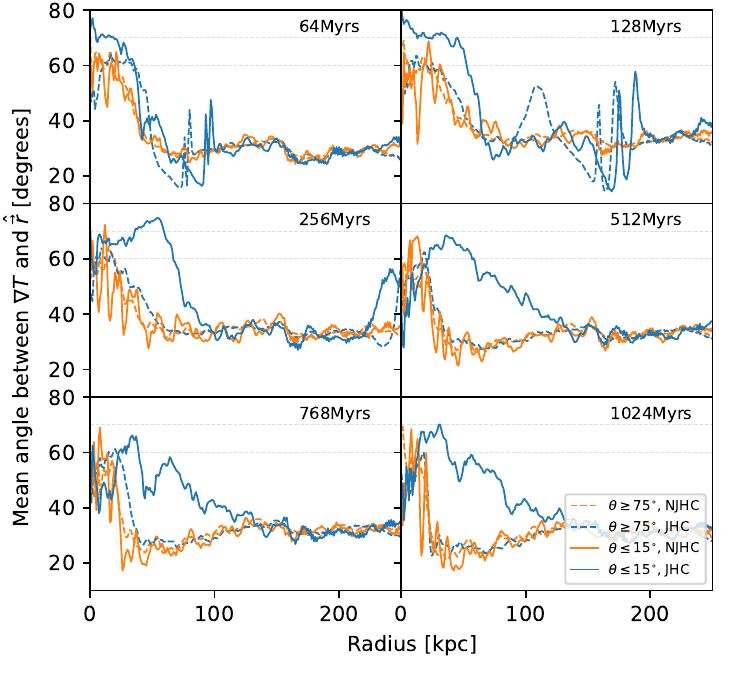}
  \caption{Mean angle of temperature gradient relative to the radius vector (large angles imply gradients mostly tangential to $\hat\vec{{r}}$), measured in a $\theta \leq 15^{\circ}$ cone around the mean jet axis (solid lines) and in the ``ruff'' region $\theta \geq 75^{\circ}$ (dashed lines) for the JHC (blue), compared to the NJHC simulation (orange) at different times.  The figure shows the clear uplift of cold gas by the jet and the resultingly more tangential temperature gradient.}
  \label{fig:t_angle}
\end{figure}

We can break this trend apart into the two contributing effects: Figure~\ref{fig:b_angle} shows the mean angle between the magnetic field vector and the radius vector, highlighting the radial field alignment within the jet cone. Then, in Fig.~\ref{fig:t_angle}, we show the mean angle between the temperature gradient and the radius vector, showing a more tangential temperature gradient around the uplifted gas in the JHC simulation. These effects are not observed in the ``ruff'' region of the JHC simulation, confirming that the jet is inducing this effect upon the system.

\subsection{The Synergistic Interplay of Jet Feedback and Conduction}

While our findings indicate that jet-enhanced thermal conduction in cool core clusters is likely marginal (see \S\ref{sec:caveats} for a discussion of caveats regarding this finding) our simulations reveal an interesting interplay between thermal conduction and jet feedback nonetheless:

Our simulations indicate that jet feedback is most efficient at increasing the entropy in the very core of the cluster, partly by uplift of low entropy gas, but also other means of heating, such as shock heating and turbulence. In fact, the large scale G-mode oscillations excited by the jet persist for well over a Gyr, driving large scale eddies and leading to increased temperatures in the cluster core, as can be seen in Fig.~\ref{fig:tempslices}.

They also reaffirm the fact that thermal conduction primarily heats the transition region of the cluster, i.e., heating the cluster from the outside in. This suggests a synergistic role played by both, which is borne out by the temperature and entropy profiles plotted in Figs.~\ref{fig:temp_ratio} and \ref{fig:frac_entropy}, showing the increase in temperature and entropy across the entire cool core when both effects act together. 

Clearly, the extent of jet heating and the magnitude of heat deposition by thermal conduction will depend on the parameter choices for both. It is beyond the scope of this paper to address these questions more exhaustively, but the work presented here encourages further investigations that explore a wider envelope of jet powers, duty cycles, and the role that thermal conduction can play in concert with jet activity.

\subsection{Caveats}
\label{sec:caveats}

The results presented above suggest that, under conservative assumptions, the interplay between jet-induced uplift and thermal conduction leads to only modest additional heating in cool core clusters, while the efficiency of the heat pump can be close to maximal if conduction operates close to the Spitzer value or is isotropic (e.g., through the action of turbulent conduction.) However, as with all numerical studies, our work is subject to limitations we will discuss next.

\subsubsection{Jet Duty Cycle and Power}

Our numerical experiments were designed to test conditions similar to those investigated in \citet{Chen_2019} at face value. The most significant limitation in this context is that we investigate the same single-episode injection at $P_{\rm jet}=10^{45}\,{\rm ergs/s}$ for a duration of 10Myrs. It is straight forward to see from Figs.~\ref{fig:tempslices} and \ref{fig:heatphase} that the uplift of the lowest entropy gas induced by such an episode is limited to approximately 100~kpc, which is still within the transition region from the cool core to the hot atmosphere.

A longer duration jet episode and/or a more powerful jet would presumably lead to larger uplift, which has the potential to substantially increase the amount of heat transfer for three main reasons:
\begin{itemize}
    \item{It would increase the temperature contrast and thus conductive potential by exposing the uplifted gas to higher temperatures at higher elevations}
    \item{It would reduce the temperature of the uplifted gas by adiabatic expansion (as outlined in \citet{Chen_2019}) and thus once again increase the temperature contrast}
    \item{It would increase the buoyancy time and thus the duration over which conduction can operate.}
\end{itemize}

The limitation to a single episode of jet activity investigated in our experiments is also a clear simplification. Repeated jet episodes along different axes may further open multiple channels for uplift and, to lowest order, we might expect the total energy injected by the heat pump to be proportional to the total jet energy injected into the cluster. However, it is also plausible that repeated jet injections in the same direction either enhance or disrupt the uplift and subsequent heating of low entropy gas. Clearly, future work should address the effects of longer jet activity and multiple episodes of jet activity either in the same direction or along different mean jet axes.

\subsubsection{Resolution and Magnetic Topology}

Like all numerical studies, our results are limited by the numerical resolution achievable with the resources available to us. The most important aspect of how resolution will affect our results is by limiting the ability to model the magnetic topology to small scales. If the magnetic field structure in the ICM is such that fields are strongly tangled below our resolution scale, our simulations will not be able to model its effect on anisotropic conduction. In that case, we would expect conduction rates to be further decreased.

The ultimate effect of resolution on how conduction is modeled in our experiments depends on the power spectrum of magnetic fluctuations we impose as initial conditions. Because we impose a magnetic topology on the ICM field that is resolved by our numerical box, we should not expect increased resolution to significantly alter the resulting conduction rates in the cool core of the cluster, where resolution in our simulations is highest. However, one may plausibly expect fluctuations of the ICM field to continue to smaller scales than those modeled in our power spectrum. 

We present as modest convergence test in \S\ref{sec:appendix} to address the question of how robust our results are against changes in resolution and find that the main conclusions appear converged. Nonetheless, future simulations at higher resolution are highly desirable and would be able to test how sensitive our results are to such spectral changes.

\subsubsection{Modeling Conduction}

The dependence of our results on the rate of conduction is an important factor in a study like this, as the maximum amount of heat deposited within an uplift time will change given a higher or lower conduction rate. To address this, we tested a very low rate of conduction, or a Spitzer-Braginskii value of $0.01$ in simulations JLC and NJLC. Shown in Fig.~\ref{fig:jet_heat}, the strongly suppressed $f_{\rm sp} = 0.01$ models show a smaller amount of total heat injected, which is consistent with this narrative.

This range of values in $f_{\rm sp}$ represents the conditions discussed in \citep{Chen_2019}. However, the assumption that $f_{\rm sp}\leq 0.1$ may be overly pessimistic. To this end, we conducted a simulation with $f_{\rm sp}\sim 1$, the largest plausible value. However, the sensitive dependence on the Courant time step on the conduction rate does not allow us to complete a suite of experiments at the same resolution as our other runs, as it would require a tenfold increase in total computing time. We thus limited the maximum refinement step to one level below that used in all other simulations. As we discuss in \S\ref{sec:appendix}, our main results are insensitive to this change in resolution. With this proviso, the simulation ran only a factor of ten slower than the fiducial run. We completed this simulation to about half the length of all other simulations and included it in the analysis shown in Fig.~\ref{fig:heat_pump}.

It is important to note that increasing or decreasing $f_{\rm sp}$ will not change the {\em relative} importance of heat conduction into the uplifted gas compared to the {\em overall} conductive heating of the cool core, as both have the same dependence on $f_{\rm sp}$. Thus, our conclusion that the excess conduction into the {\em uplifted} gas is at best a modest fraction of the {\em overall} heat deposited into the cool core by conduction (regardless of the presence or absence of jets) should be robust against changes in $f_{\rm sp}$.

Another effect that calls for further investigation is the possibility of some amount of cross-field, or isotropic, conduction allowing for some degree of non-magnetically suppressed conduction, predominantly through the effects of turbulent conduction. This would create the possibility for additional heating on top of our calculated heat injection. We explored the effect of isotropic conduction to a limited extent in our JHIC and NJHIC runs. We find that, on average, {\em anisotropic} conduction is suppressed by a factor of 5 to 7 relative to {\em isotropic} conduction.

ICM turbulence not resolved in our simulations (driven by the action of the jets, motions of galaxies in the cluster, cold fronts, or minor mergers) may lead to signficantly increased turbulent conduction rates \citep{ruszkowski_2010, Lazarian_2020}. We do not include any implicit modeling for turbulent conduction beyond any turbulence present in our simulations. In order to estimate the relative importance of what we might expect for the overall conduction rates, we can compare the Spitzer-Braginskii rate from eq.~(\ref{eq:spitzer}) with the turbulent conduction:
\begin{align}
\kappa_{\mathrm{turb}} 
&\sim v_{\mathrm{turb}}\,\ell_{\mathrm{drive}} \nonumber\\
&\sim 5\times10^{29}\,\mathrm{cm^{2}\,s^{-1}}
\left(\frac{\ell_{\mathrm{drive}}}{10\,\mathrm{kpc}}\right)
\left(\frac{v_{\mathrm{turb}}}{150\,\mathrm{km\,s^{-1}}}\right)
\end{align}
Where $v_{\rm turb}$ is the turbulent velocity and $\ell_{\rm drive}$ is the driving scale of the turbulence. Comparing this to the classic Spitzer-Braginskii value from eq.~(\ref{eq:spitzer}), we find
\begin{equation}
    \frac{\kappa_{\rm turb}}{\kappa_{\rm sp}} \sim 0.2
\end{equation}
which suggests that turbulent and microscopic thermal conduction may be of the same order of magnitude (depending on the detailed parameter choices). Thus, depending on the efficacy of turbulent conduction, the heat pump efficiencies estimated from our JHC and JLC runs may underestimate the true value, once again calling for further investigation.

The most straight forward way to model turbulent conduction in future experiments would be to include an isotropic, temperature-independent conduction coefficient to operate in alongside the anisotropic conduction modeled in our numerical experiments. Since turbulent conduction can be expected to be quasi-isotropic, our JSC run might even underestimate the effects of conduction in such a case. However, an additional isotropic run close to $f_{\rm sp}$ would have been beyond the computational resources available for this study and remains to be explored in future work.

\section{Conclusions}

We conducted a series of numerical experiments to evaluate the interplay of jet feedback and thermal conduction in the cool core clusters. The aim was to test whether large feedback efficiency gains through the so-called heat pump effect first discussed in \citet{Chen_2019} may be feasible even if thermal conduction is strongly suppressed below the Spitzer-Braginskii value. The heat pump effect is the result of dynamical uplift of low entropy gas by the action of jets, which can then be brought into thermal contact with the heat bath of hot cluster gas at larger radii. 

To this end, we implemented anisotropic conduction, randomized initial magnetic field configuration, and scalar tracer variables to explicitly track the conductive deposition of heat into the \texttt{FLASH} code.

Our findings are as follows:
\begin{enumerate}

\item{We find that the action of jets induces long term buoyancy oscillations which lead to several episodes of uplift and downdraft of low entropy gas.}
\item{Under the conditions discussed in \citet{Chen_2019}, both anisotropic thermal conduction and jet feedback lead to an increase in the central entropy and temperature of the cluster.}
\item{A single episode of jet injection with a total energy of $3\times 10^{59}\,{\rm ergs}$ by itself raises the entropy in the central 20\,{\rm kpc} zone by approximately 12\%, while the temperature increase by approximate 6\%.}
\item{Anisotropic conduction alone, operating at 10\% of the Spitzer-Braginskii rate and acting for the full Gyr duration of the simulation, raises the entropy in the cool core transition zone between 30~kpc and 120~kpc by up to 4\% and the temperature by up to 2\%. These small gains are expected to be directly proportional to time and conduction coefficient.}
\item{In simulations with both jet feedback and anisotropic conduction suppressed at or below 10\% of the classic Spitzer-Braginskii rate, we find that the resulting increases in entropy and temperature are consistent with the sum of the increases derived from simulations of jet feedback only and conduction only, indicating that any heat pump effect present is small. We find that, for an uplift of 100~kpc and conduction rate at 10\% of the Spitzer-Braginskii value, the total heat deposited into the uplifted gas is of order $\Delta H_{\rm lift} \sim 1-2\times 10^{58}\,{\rm ergs}$.}
\item{We find that for more optimistic assumptions, such as anisotropic conduction at the classic Spitzer-Braginskii value or for isotropic conduction at or above $f_{\rm sp} \gtrsim 0.1$, the heat pump deposits up to $H_{\rm lift} \sim 5\times 10^{58}\,{\rm ergs}$ into the uplifted gas.}
\item{Comparing the heat deposited into the uplifted gas to the maximum possible amount of heat available for heat transfer $H_{\rm max}$, given the temperature differential of the uplifted gas to the equilibrium temperature, we find that anisotropic conduction at the Spitzer rate reaches approximately 50\% of the maximum possible value, and an overall heat pump efficiency of $\eta_{\rm max}\sim 80\%$. The same is true for isotropic conduction at a suppression factor of $f_{\rm sp} \gtrsim 0.1$. For anisotropic conduction with strong suppression below the Spitzer-Braginskii rate ($f_{\rm sp}\ll 1$) we find that the heat pump efficiency is well approximated by $\eta_{\rm fp \ll 1} \sim f_{\rm sp}\eta_{\rm max}$.}
\item{The lack of a strong heat pump effect for $f_{\rm sp} \lesssim 0.1$ can be readily understood as the result of the self-limiting character of this type of uplift-driven heat pump: Magnetic draping by the uplift of low entropy gas results in magnetic field lines predominantly oriented perpendicular to the lateral temperature gradient created by the uplift, significantly suppressing conductive heating of the uplifted gas.}
\item{Regardless of whether additional heating by the heat pump mechanism occurs, we find that jet feedback and thermal conduction complement each other in that jet feedback can effectively lift the temperature of the very inner core of the cluster, while conduction is most effective at increasing the temperature of the transition region between cool core and hot atmosphere.}
\item{A longer jet duty cycle or larger power, leading to larger uplift, may lead to an increased heat pump efficiency and calls for future simulations that extend this work beyond the setup studied in \citet{Chen_2019} and expanded upon here.\\}
\end{enumerate}

This study lays the groundwork for a variety of follow-up investigations, including running more models to account for and test the above-mentioned effects of larger jet duty cycles, conduction prescriptions, long term precession of the jet axis, and the inclusion of additional cluster physics.

\begin{acknowledgements}
The authors acknowledge Access/XSEDE, NASA's NAS, and Wisconsin Institute for Discovery's CHTC for providing the computing resources needed to conduct these studies. Support was also provided by the Graduate School and the Office of the Vice Chancellor for Research at the University of Wisconsin-Madison with funding from the Wisconsin Alumni Research Foundation. J.S.~acknowledges that this material is based upon work supported by the National Science Foundation Graduate Research Fellowship Program under Grant No. DGE-1747503. Any opinions, findings, and conclusions or recommendations expressed in this material are those of the author(s) and do not necessarily reflect the views of the National Science Foundation. S.H.~acknowledges support from NASA grant 80NSSC23K0014 and NASA/Chandra grant TM2-23007X. M.R.~acknowledges support from the National Science Foundation Collaborative Research Grant NSF AST-2009227.

\end{acknowledgements}


\bibliographystyle{aasjournal}
\bibliography{bibliography}

\appendix{}
\section{Convergence}
\label{sec:appendix}
\begin{figure}[t]
  \centering
  \includegraphics[width=.435\columnwidth]{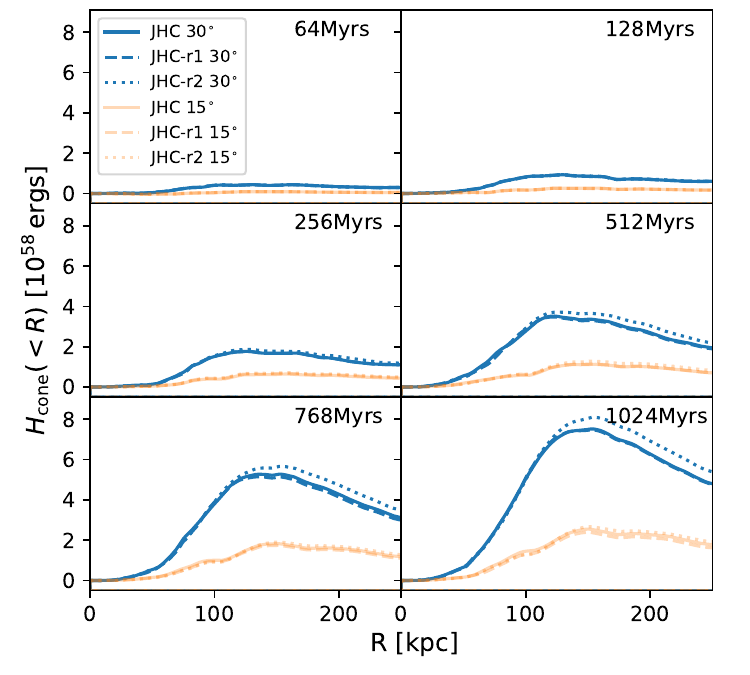}
  \includegraphics[width=.545\columnwidth]{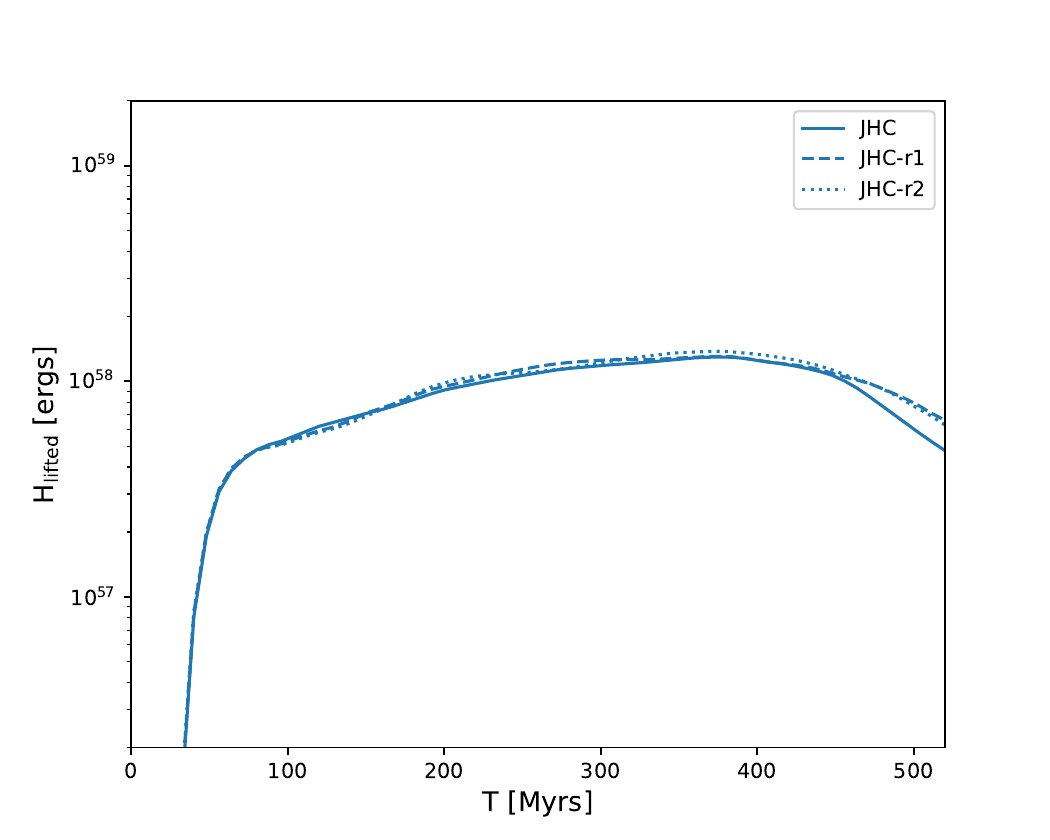}    
  \caption{Convergence study comparison for fiducial JHC run and two lower resolution criteria, denoted as JHC-r1 and JHC-r2, plotted as dashed and dotted lines, respectively. {\em Left panel:} Resolution comparison for Fig.~\ref{fig:cone}, showing the heat $H(<R)$ accumulated within radius $R$ in cones of half opening angles of $15^{\circ}$ and $30^{\circ}$ in blue and orange, respectively. {\em Right panel:} Resolution comparison for Fig.~\ref{fig:heat_pump}, showing the same quantity (total heat deposited into uplifted gas) for three different AMR resolution limits.}
  \label{fig:jet_heat_res}
\end{figure}\textbf{}
To verify the robustness of our results against resolution dependence, we performed a small convergence calculation for our fiducial JHC simulation by rerunning the simulation with lower resolution. We changed the refinement criteria for these simulations in two different ways. First, we simply reduced the maximum refinement level of the simulation by one level. We denote this simulation as run JHC-r1. This is the prescription used to make the JSC run computationally feasible. 

Second, we reduced the refinement of all blocks by one level, which represents, a true reduction in resolution by a factor of two. We denote this run as JHC-r2. We plot the different runs---JHC, JHC-r1, JHC-r2---as solid, dashed, and dotted lines, respectively.

We then calculated the core diagnostic quantities presented in \S\ref{sec:discussion} for these two runs and show them in Fig.~\ref{fig:jet_heat_res}. The left panel of that figure shows the equivalent diagnostic from Fig.~\ref{fig:jet_heat}. The figures show reasonable convergence.

The right panel of Fig.~\ref{fig:jet_heat_res} shows the equivalent diagnostic from Fig.~\ref{fig:heat_pump}, once again showing reasonable convergence. While we see a deviation starts appearing towards the late times on this figure, we note that, at those times, the bulk of the uplifted material sinks below the mask and large fluctuations appear between simulations. The key quantity we measure--the maximum of the curves--is converged.

We conclude that, within the relatively large levels of uncertainty imposed by the wide range of plausible parameter choices, our results are robust against resolution changes.

\end{document}